\documentclass[aps,prd,preprintnumbers,showpacs,nofootinbib,floatfix]{revtex4}
\usepackage{amsmath,amssymb,amsbsy,mathbbol}
\usepackage{graphicx,subfigure,psfrag}

\def\Tr{\mathrm{Tr}}
\def\tilde{\widetilde}

\begin{document}

\preprint{UW-PT 04-20, SHEP-0438}

\title{Twisted mass chiral perturbation theory at next-to-leading order}

\author{Stephen R. Sharpe}
\email[]{sharpe@phys.washington.edu}
\altaffiliation[Permanent address: ]{
Physics Department, University of Washington,
  Seattle, WA 98195-1560, USA}
\affiliation{School of Physics and Astronomy, University of Southampton,
Southampton, SO17 1BJ, UK}

\author{Jackson M. S. Wu}
\email[]{jmw@phys.washington.edu}
\affiliation{Physics Department, University of Washington,
  Seattle, WA 98195-1560, USA}

\date{January 21, 2005}

\begin{abstract}
We study the properties of pions in twisted
mass lattice QCD (with two degenerate flavors) 
using chiral perturbation theory ($\chi$PT).
We work to next-to-leading order (NLO)
in a power counting scheme in which
$m_q \sim a \Lambda_{\rm QCD}^2$, with $m_q$ the physical quark
mass and $a$ the lattice spacing. 
We argue that automatic $O(a)$ improvement
of physical quantities at maximal twist, which has been demonstrated
in general if $m_q \gg a \Lambda_{\rm QCD}^2$,
holds even if $m_q \sim a \Lambda_{\rm QCD}^2$, 
as long as one uses an appropriate
non-perturbative definition of the twist angle,
with the caveat that we have shown this only through
NLO in our chiral expansion.
We demonstrate this with explicit calculations, for arbitrary
twist angle, of all pionic quantities that involve no more
than a single pion in the initial and final states: masses,
decay constants, form factors and condensates, as well as
the differences between alternate definitions of twist angle. 
We also calculate the axial and pseudoscalar
form factors of the pion, quantities which violate flavor and parity,
and which vanish in the continuum limit. These are of interest
because they are not automatically $O(a)$ improved at maximal twist.
They allow a determination of the unknown low energy constants 
introduced by discretization errors, and provide tests of the accuracy
of $\chi$PT at NLO.
We extend our results into the regime where
$m_q \sim a^2 \Lambda_{\rm QCD}^3$, and argue in favor
of a recent proposal that automatic $O(a)$ improvement
at maximal twist remains valid in this regime.
\end{abstract}


\maketitle

\section{\label{sec:intro} Introduction}

Twisted mass lattice QCD (tmLQCD)~\cite{FetalLatt99,Fetal01} is an
alternative regularization for lattice QCD that has recently received
considerable attention.\footnote{For a recent review see
Ref.~\cite{FrezLatt04}.} 
It has the potential to match the attractive
features of improved staggered fermions (efficient
simulations~\cite{Kenn04}, absence of ``exceptional
configurations''~\cite{FetalLatt99}, $O(a)$ improvement at maximal
twist~\cite{FR03}, operator mixing as in the
continuum~\cite{Fetal01,Pena04,FR04}) while not sharing the
disadvantage of needing to take roots of the determinant to remove
unwanted degrees of freedom. Initial numerical investigations of the
unquenched theory~\cite{Faretal04,FarLatt04} have, however, found
surprisingly large discretization errors, which manifest themselves as
a non-trivial phase structure.\footnote{%
  Indications of surprisingly
  large discretization errors have also been found in quenched tmLQCD
  at maximal twist~\cite{Biet04}.
  On the other hand, use of an improved gauge action
  reduces the discretization errors~\cite{Faretal04B}.}
While these errors do not pose a
fundamental problem, and, indeed, appear to conform to the
expectations of chiral perturbation theory applied to the lattice
theory~\cite{SS98,Mun04,Scor04,SW04}, they do suggest that a thorough
investigation of the impact of discretization errors on the theory is
called for. 

Such an investigation is possible by applying the methods of chiral
perturbation theory to tmLQCD at non-zero lattice spacing. The
discretization errors are included systematically in a joint expansion
in the lattice spacing, $a$, and the quark mass, $m_q$. The resulting
``twisted mass chiral perturbation theory'' (tm$\chi$PT) has been
formulated previously~\cite{MSch04,Scor04,SW04}, 
building on earlier work for the untwisted Wilson 
theory~\cite{SS98,RS02,BRS03}.
It has been used to study pion masses and decay constants
in the regime where $m_q\gg a \Lambda_{QCD}^2$~\cite{MSch04},
and to study the phase structure of tmLQCD when 
$m_q\sim a^2 \Lambda_{QCD}^3$~\cite{Mun04,MunLatt04,Scor04,
SW04,SWLatt04,AB04}.
Here we build upon our previous work~\cite{SW04},
in which we determined the chiral Lagrangian at next-to-leading order (NLO)
in a power counting scheme in which we treat $m_q \sim a \Lambda_{QCD}^2$.
We use this to
study all the quantities involving pions
that do not involve final state interactions and are
thus straightforward to calculable in simulations:
masses (previously calculated in this regime in Ref.~\cite{Scor04}),
 condensates, vacuum to pion matrix elements, and matrix
elements between single pion states (i.e. form factors). 
The operators we use are the
vector and axial currents and the scalar and pseudoscalar densities. 

Calculating these quantities allows us to address some of the issues
that are of concern when using tmLQCD. In particular, what is the
impact of the breaking of flavor and parity symmetries at non-zero
lattice spacing? What are good quantities to use to determine the
typical size of discretization errors? How does the automatic $O(a)$
improvement at maximal twist manifest itself in tm$\chi$PT? How do
different definitions of the twist angle, all of which are
equivalent in the continuum limit, differ at non-zero lattice spacing?
How can we test the reliability of tm$\chi$PT applied at a given
order? We return to these questions in the concluding section.

Another issue, which is of particular importance for practical simulations,
is the condition
on the quark mass that must be enforced in order that 
physical quantities are automatically $O(a)$ improved at maximal twist.
Does one need to enforce (A) $m_q \gg a \Lambda_{\rm QCD}^2$,
(B) $m_q \sim a \Lambda_{\rm QCD}^2 \gg a^2 \Lambda_{\rm QCD}^3$,
or (C) $m_q \ge c a^2 \Lambda_{\rm QCD}^3$, 
with $c$ a determinable constant of $O(1)$?
If one takes $a^{-1}=2\;$GeV and $\Lambda_{\rm QCD}=0.3\;$GeV,
these conditions become (A) $m_q \gg 45\;$MeV, (B) $m_q \gg 7\;$MeV,
and (C) $m_q \gtrsim 7\;$MeV, respectively.
Since the average light quark mass is $\sim 3$ MeV
(at a renormalization scale of $2\;$GeV---for a recent
review see Ref.~\cite{Hash04}), we need to
be able to use condition (B) if lattice masses are to allow
extrapolation to physical values, 
or condition (C) if they are to approach them.
Condition (A) is very restrictive.

On the theoretical side, Refs.~\cite{FR03,FrezLatt04} argue that, 
in general, one must use the strongest condition, (A), 
although it is possible that the intermediate condition, (B),
can be used if one is dealing with $O(a)$ improved quantities,
e.g. if using an $O(a)$ improved quark action.
By contrast, it has been recently proposed
in Ref.~\cite{AB04} that there is no lower limit on
$m_q$ in order for automatic $O(a)$ improvement to apply,
as long as one uses an appropriate definition of twist angle.

We will argue here in favor of the weakest condition, (C),
with the constant $c$ either of $O(1)$, or vanishing,
depending on the sign of an unknown constant
in the chiral Lagrangian.
Our arguments are based on tm$\chi$PT applied in the regimes
where conditions (B) and (C) hold. The former we call
the ``generic small mass'' (GSM) regime, 
the latter the ``Aoki'' regime (since
it is in this region that non-trivial phase
structure due to discretization effects appears~\cite{Aoki}).
For most of the paper we consider the GSM regime, in which
we expect the bulk of future simulations to be done.
We give a general argument for automatic $O(a)$ improvement
in this regime (which requires the use of an appropriate
non-perturbative definition of twist angle),
and support it with results for the physical quantities
listed above.
We then extend the results to the Aoki regime, where, 
working to leading order (LO) in
an expansion in which $m_q \sim a^2 \Lambda_{\rm QCD}^3$,
we argue that condition (B) can be relaxed to (C).

We stress that our discussion of $O(a)$ improvement is within the
context of chiral perturbation theory up to a given order (NLO in the
GSM regime, LO in the Aoki regime). This means that we only control those
$O(a)$ corrections that are expected to be dominant in the regimes under study.
For example, in both regimes we control corrections
to the pion decay constant, $f_\pi$, which are of relative size $a \Lambda_{\rm QCD}$,
but do not control those of relative size $a m_q$.
The latter corrections arise from terms in the chiral Lagrangian of
$O(a^3)$ and $O(a^4)$ in the GSM and Aoki regimes, respectively, which are,
in both cases, of higher order than we work.

The remainder of this article is organized as follows. In the next
section, we recall the definition of tmLQCD and the construction
of the corresponding NLO chiral Lagrangian, and use this to discuss
the condition on the quark mass needed to obtain automatic
$O(a)$ improvement at maximal twist.
In Sec.~\ref{sec:GSM} we use the Lagrangian, including sources
for currents and densities, to determine the twist angle
non-perturbatively, and to calculate the 
physical quantities listed above. In the final subsection,
we give a summary of our results 
and a suggestion for how to use them in practice.
In Sec.~\ref{sec:Aoki} we show how our results extend into the Aoki
regime, and comment on the considerations of Ref.~\cite{AB04}. We
conclude in Sec.~\ref{sec:conc}, and suggest various avenues for
future work.  

Some preliminary results from this paper have been presented
in Ref.~\cite{SW04}.

\section{\label{sec:EChiL} The effective chiral Lagrangian}

The theory we consider is tmLQCD with a degenerate doublet
of quarks. The fermionic part of the Euclidean lattice action of the
theory in the so-called ``twisted basis'' is~\cite{Fetal01,FR03}: 
\begin{align} \label{E:action}
S^L_F =& \; \sum_{x} \bar{\psi_l}(x)
\Big[\frac{1}{2} \sum_{\mu} \gamma_\mu (\nabla^\star_\mu + \nabla_\mu)
- \frac{r}{2} \sum_{\mu} \nabla^\star_\mu \nabla_\mu 
+ m_0 + i \gamma_5 \tau_3 \mu_0 \Big] \psi_l(x),
\end{align}
where $\psi_l$ and $\bar\psi_l$ are the dimensionless bare lattice fields (with
``$l$" standing for lattice and not indicating left-handed), and
$\nabla_\mu$ and $\nabla^\star_\mu$ are the usual covariant forward
and backward dimensionless lattice derivatives, respectively.
The field $\psi_l$ is a flavor doublet, and $\tau_3$ acts in flavor
space and is normalized so that $\tau_3^2=1$. The bare normal mass,
$m_0$, and the bare twisted mass, $\mu_0$, both of which are dimensionless,
are taken to be
proportional to the identity matrix in flavor space. 

The effective continuum chiral theory is derived using the two-step
procedure of Ref.~\cite{SS98}. In Ref.~\cite{SW04},
we have carried this procedure
out to NLO in an expansion in which we treat quark masses and the
leading discretization errors as symmetry breaking parameters
of the same size.
Here we recall only the essential details.

Following the program of Symanzik~\cite{Sym83}, we first write down an
effective continuum Lagrangian at the quark level which describes the
long distance physics of the underlying lattice theory. Its form is
constrained by the symmetries of the lattice theory to be
\begin{align} \label{E:CLeff}
\mathcal{L}_{\rm eff} &= \mathcal{L}_g + \bar{\psi}
(D \!\!\!\!/ + m + i \gamma_5 \tau_3 \mu) {\psi} 
+ b_1 a \bar{\psi} i \sigma_{\mu\nu} F_{\mu\nu} \psi 
+ O(a^2) \,.
\end{align}
Here $\mathcal{L}_g$ is the continuum gluon Lagrangian, $m$ is the physical
quark mass, defined in the usual way by 
\begin{equation} 
m = Z_m(m_0 - \widetilde m_c)/a \,,
\end{equation}
and $\mu$ is the physical twisted mass 
\begin{equation}
\mu = Z_\mu \mu_0/a = Z_P^{-1} \mu_0/a \,,
\end{equation}
with $Z_P$ the matching factor for the pseudoscalar density.
The quantity $\tilde m_c$ is the critical mass, aside from
an $O(a)$ shift. This shift, and methods for determining
the critical mass after its inclusion, are discussed below. Note
that the lattice symmetries forbid additive renormalization of
$\mu_0$~\cite{Fetal01}. 
Although the $O(a^2)$ terms in Eq.~(\ref{E:CLeff})
are of an appropriate size to be included
at NLO, they do not break the continuum symmetries any further than
the terms explicitly shown, and thus do not lead to any additional
operators in the effective chiral theory~\cite{SW04}. 
Thus we do not need their explicit form.

We reiterate here that we have dropped from ${\cal L}_{\rm eff}$ terms
proportional to $a \mu^2 \bar\psi \psi$, $a m^2 \bar\psi\psi$ and
$a \mu m \bar\psi i\gamma_5 \tau_3 \psi$. These terms are allowed by
the lattice symmetries~\cite{FSW01}, and do lead to corrections linear in $a$,
but are of next-to-next-to-leading order (NNLO) in our power counting.

In the continuum limit
[where the $b_1$ and higher order terms in (\ref{E:CLeff}) vanish] 
the apparent flavor-parity breaking due to $\mu$ is misleading, since
it can be rotated away by a non-anomalous axial rotation. Thus tmLQCD
is, in this limit, equivalent to QCD with two degenerate quarks
of mass $\sqrt{m^2 + \mu^2}$. This has been established in detail
in Ref.~\cite{Fetal01}.
In this paper, we work away from the continuum limit, taking
$m \sim \mu \sim p^2 \sim a \Lambda_{\rm QCD}^2$. Thus the symmetry
breaking induced by the $b_1$ term will play a crucial role.

The next step is to match the continuum effective 
Lagrangian (\ref{E:CLeff}) onto a generalized chiral Lagrangian.
Working at NLO in our power counting, we found~\cite{SW04}
(after simplifying using the properties of $SU(2)$ matrices)
\begin{align} \label{E:ChiJLeff}
\mathcal{L}_\chi &= 
 \frac{f^2}{4} \mathrm{Tr}(D_\mu \Sigma D_\mu \Sigma^\dagger)
-\frac{f^2}{4} \mathrm{Tr}(\chi^{\dagger} \Sigma + \Sigma^\dagger\chi) 
-\frac{f^2}{4} \mathrm{Tr}(\hat{A}^{\dagger} \Sigma + 
               \Sigma^\dagger\hat{A}) \notag \\ 
&\quad
- L_1 \mathrm{Tr}(D_\mu \Sigma D_\mu \Sigma^\dagger)^2
- L_2 \mathrm{Tr}(D_\mu \Sigma D_\nu \Sigma^\dagger)
      \mathrm{Tr}(D_\mu \Sigma D_\nu \Sigma^\dagger) \notag \\
&\quad 
+ L_{45} \mathrm{Tr}(D_\mu \Sigma^\dagger D_\mu \Sigma)
         \mathrm{Tr}(\chi^{\dagger} \Sigma +  \Sigma^\dagger\chi)
\notag \\ &\quad
+ L_5 \left\{
\mathrm{Tr}\left[(D_\mu \Sigma^\dagger D_\mu \Sigma)
      (\chi^{\dagger} \Sigma +  \Sigma^\dagger\chi)\right]
-\mathrm{Tr}(D_\mu \Sigma^\dagger D_\mu \Sigma)
      \mathrm{Tr}(\chi^{\dagger} \Sigma +  \Sigma^\dagger\chi)/2 \right\}
\notag \\ &\quad
- L_{68} 
\big[\mathrm{Tr}(\chi^{\dagger} \Sigma + \Sigma^\dagger\chi)\big]^2
- L_8 \left\{
\mathrm{Tr}\big[(\chi^{\dagger} \Sigma + \Sigma^\dagger\chi)^2\big]
- \big[\mathrm{Tr}(\chi^{\dagger} \Sigma + \Sigma^\dagger\chi)\big]^2/2
\right\}
\notag \\ &\quad
- L_7 \big[\mathrm{Tr}(\chi^{\dagger} \Sigma - \Sigma^\dagger\chi)\big]^2
\notag \\ &\quad
+ i L_9 \mathrm{Tr}(L_{\mu\nu} D_\mu \Sigma D_\nu \Sigma^\dagger +
           R_{\mu\nu} D_\mu \Sigma^\dagger D_\nu \Sigma)
+ L_{10} \mathrm{Tr}(L_{\mu\nu} \Sigma R_{\mu\nu} \Sigma) \notag \\
&\quad
+ W_{45} \mathrm{Tr}(D_\mu \Sigma^\dagger D_\mu \Sigma)
         \mathrm{Tr}(\hat{A}^{\dagger} \Sigma +
         \Sigma^{\dagger}\hat{A})
- W_{68} \mathrm{Tr}(\chi^{\dagger} \Sigma + \Sigma^\dagger\chi) 
         \mathrm{Tr}(\hat{A}^{\dagger} \Sigma + \Sigma^{\dagger}\hat{A})
\notag \\
&\quad
- W_{68}' \big[\mathrm{Tr}(\hat{A}^{\dagger} \Sigma +
       \Sigma^{\dagger}\hat{A})\big]^2
+ W_{10} \mathrm{Tr}(D_\mu \hat{A}^\dagger D_\mu \Sigma + 
          D_\mu \Sigma^\dagger D_\mu \hat{A}) \notag \\
&\quad + H_1\mathrm{Tr}(L_{\mu\nu}L_{\mu\nu}+R_{\mu\nu}R_{\mu\nu})
- H_2 \mathrm{Tr}(\chi^\dagger\chi)
- H'_2 \mathrm{Tr}(\hat{A}^\dagger \chi+ \chi^\dagger \hat{A})
- H_3 \mathrm{Tr}(\hat{A}^\dagger \hat{A}) \,.
\end{align}
As usual, the field $\Sigma$ is $SU(2)$ matrix-valued, and transforms under
the chiral group $SU(2)_L \times SU(2)_R$ as
\begin{equation}
\Sigma \rightarrow L \Sigma R^{\dagger}, \qquad 
L \in SU(2)_L, \; R \in SU(2)_R. 
\end{equation}
The quantities $\chi$ and $\hat{A}$ are spurions for the
quark masses and discretization errors, respectively.
At the end of the analysis they are set to the constant values
\begin{equation}
\chi \longrightarrow 2 B_0 (m + i \tau^3 \mu) 
\equiv \hat{m} + i \tau^3 \hat{\mu}, \qquad\qquad 
\hat A \longrightarrow 2 W_0 \, a \equiv \hat{a} \,,
\end{equation}
where $B_0$ and $W_0$ are unknown dimensionful constants,
and we have defined the quantities $\hat{m}$, $\hat{\mu}$ and
$\hat{a}$. 
We have included sources for left- and right-handed currents
in the standard way using covariant derivatives and
associated field strengths, {\em e.g.}
\begin{equation}
D_\mu \Sigma = \partial_\mu \Sigma 
- i \ell_\mu \Sigma + i \Sigma r_\mu\,,
\qquad
L_{\mu\nu} = \partial_\mu \ell_\nu - \partial_\nu \ell_\mu 
+ i [\ell_\mu,\ell_\nu]\,.
\end{equation}
Sources for the scalar and pseudoscalar densities are similarly
included by writing $\chi = 2 B_0 (s + i p)$.
Note that $\ell_\mu$, $r_\mu$, $s$ and $p$ are all
hermitian matrix fields.
Finally, we note that at NLO we should also include the
Wess-Zumino-Witten term~\cite{WZW}. We do not, however, consider here
any processes to which it contributes (e.g. $\pi^0\to\gamma\gamma$), 
and so do not write it out explicitly.

The chiral Lagrangian contains a number of parameters
which are not determined by symmetries---the
low energy constants (LECs). At leading order there are
the usual $f$ (normalized so that $f_\pi=93\;$MeV)
and $B_0$, supplemented by the additional constant $W_0$
introduced by discretization errors~\cite{SS98}.
At NLO there are significantly more LECs. The $L_i$ are
the usual Gasser-Leutwyler constants of continuum $\chi$PT~\cite{GL84}.
We have introduced the combinations
$L_{45}=L_4+L_5/2$ and $L_{68}=L_6+L_8/2$, which are useful
for two flavors.
The $W_i$ are introduced by discretization 
errors at linear order in the lattice spacing~\cite{RS02},
while the $W'_i$ appear at quadratic order~\cite{BRS03}.
Again, certain combinations are useful:
$W_{45}=W_4+W_5/2$, $W_{68}=W_6+W_8/2$ and $W'_{68}=W'_6+W'_8/2$. 
Finally, the $H_i$ involve sources alone, and give rise
to contact terms in correlation functions of currents and
densities, as well as contributing to their vacuum expectation
values.\footnote{%
Strictly speaking the $H_i$ and $H'_i$ are not low-energy
constants, since they absorb short-distance divergences.
In fact, in Ref.~\cite{GL84} they are denoted ``high-energy constants''.
Nevertheless, for brevity, we refer to them also as LECs.}
 The $H'_i$ are the extra such terms introduced by
discretization errors (which we did not include explicitly
in Ref.~\cite{SW04} but will need here).

It is useful to understand the relation of our result for 
${\cal L}_\chi$ to that in Ref.~\cite{MSch04}.
In that work, the authors used a different power-counting, namely
$a \Lambda_{\rm QCD} \sim (m^2 + \mu^2)/\Lambda_{\rm QCD}^2$. 
They worked to linear order in $a$, and without sources. 
Our results agree with theirs to the order they worked.
They also changed variables
using an axial rotation so that $\chi$ was diagonal. This
has the advantage of moving the twist entirely 
into the term caused by discretization errors, 
$\hat A \longrightarrow \hat{a} (m - i \mu\tau_3)/\sqrt{m^2+\mu^2}$.
It was then straightforward to see that discretization
errors in $m_\pi$ and $f_\pi$ are  proportional to
$\hat{a} m /\sqrt{m^2+\mu^2}$ and thus vanish under
mass averaging ($m\leftrightarrow -m$) and at maximal twist
($m=0$).

In our power counting, where discretization errors are superficially
as large as quark mass effects, we must follow a different strategy
to see automatic $O(a)$ improvement. The key point, first noted
in Ref.~\cite{SS98}, is that the leading discretization effect
[the third term in ${\cal L}_\chi$ in (\ref{E:ChiJLeff})] has the
same form as the leading mass term [the second term
in ${\cal L}_\chi$]. This follows directly from the
fact that the $b_1$ term in the quark-level effective Lagrangian
(\ref{E:CLeff}) has the same chiral transformation properties as the quark
mass term. Thus the LO
chiral Lagrangian is unchanged from its continuum 
form if one uses the shifted spurion
\begin{equation}
\chi' \equiv \chi + \hat{A} \,.
\end{equation}
This corresponds at the quark level to a redefinition of
the untwisted component of the quark mass from $m$ to
\begin{equation}
\label{E:m'def}
m' \equiv m + a W_0/B_0 = (\hat{m} + \hat{a})/(2B_0) \,.
\end{equation}
This shift corresponds to an $O(a)$ correction to the critical mass,
so that it becomes
\begin{equation}
m_c=\tilde m_c - a W_0/B_0
\,.
\end{equation}
In practice one automatically
determines $m_c$ (and thus automatically uses $m'$ rather than $m$)
if one uses an appropriate non-perturbative definition of
twist angle. This is discussed in detail in the following
section and summarized in the last subsection of Sec.~\ref{sec:GSM}.

Since the LO Lagrangian takes the continuum form,
the vacuum expectation value of $\Sigma$ at this order is
that which cancels out the twist in the shifted mass matrix:
\begin{equation} \label{E:VacLO}
\langle0|\Sigma|0\rangle_{LO} \equiv \Sigma_{0} = 
\frac{\hat{m} + \hat{a} + i \hat{\mu} \tau_3}{M'}
\equiv \exp(i \omega_0 \tau_3) \,,
\end{equation}
where
\begin{equation} 
\label{E:M'def}
M' = |\chi'| = {\sqrt{(\hat{m}+\hat{a})^2 +\hat{\mu}^2}}
\end{equation}
is the LO result for the pion mass-squared. 
If we define the physical quark mass by
\begin{equation}
m_q = \sqrt{m'^2 + \mu^2}
\,,
\end{equation}
then it follows from (\ref{E:VacLO}) that
\begin{equation}
c_0 \equiv \cos\omega_0 = m'/m_q \,,\qquad
s_0 \equiv \sin\omega_0 = \mu /m_q \,.
\end{equation}
Equation (\ref{E:VacLO}) is one of the definitions of twist angle that
we will use, although it is not the most simple to
determine through simulations.
It is important to realize that it differs by $O(1)$
from the naive definition
$\tan\omega = \mu/m$
when $m\sim a\Lambda_{\rm QCD}^2$.
When working in this mass regime, it is thus crucial to
use the shifted mass $m'$, or, equivalently at NLO, one of
the definitions of twist angle discussed below.
This point has also been emphasized in Ref.~\cite{AB04}.

\bigskip

We can now present our argument for why automatic
$O(a)$ improvement holds even when $m\sim \mu\sim a\Lambda_{\rm QCD}^2$.
This will be borne out later by our detailed results.
We begin by expressing the chiral Lagrangian in terms of
$\chi'$ instead of $\chi$:
\begin{align} \label{E:ChiJLeffpr}
\mathcal{L}_\chi &= 
 \frac{f^2}{4} \mathrm{Tr}(D_\mu \Sigma D_\mu \Sigma^\dagger)
-\frac{f^2}{4} \mathrm{Tr}(\chi'^{\dagger} \Sigma + \Sigma^\dagger\chi') 
\notag \\ &\quad
- L_1 \mathrm{Tr}(D_\mu \Sigma D_\mu \Sigma^\dagger)^2
- L_2 \mathrm{Tr}(D_\mu \Sigma D_\nu \Sigma^\dagger)
      \mathrm{Tr}(D_\mu \Sigma D_\nu \Sigma^\dagger) 
\notag \\&\quad 
+ L_{45} \mathrm{Tr}(D_\mu \Sigma^\dagger D_\mu \Sigma)
         \mathrm{Tr}(\chi'^{\dagger} \Sigma +  \Sigma^\dagger\chi')
- L_{68} 
\big[\mathrm{Tr}(\chi'^{\dagger} \Sigma + \Sigma^\dagger\chi')\big]^2
\notag \\ &\quad
+ L_5 \left\{
\mathrm{Tr}\left[(D_\mu \Sigma^\dagger D_\mu \Sigma)
      (\chi'^{\dagger} \Sigma +  \Sigma^\dagger\chi')\right]
-\mathrm{Tr}(D_\mu \Sigma^\dagger D_\mu \Sigma)
      \mathrm{Tr}(\chi'^{\dagger} \Sigma +  \Sigma^\dagger\chi')/2 \right\}
\notag \\ &\quad
- L_7 \big[\mathrm{Tr}(\chi'^{\dagger} \Sigma - \Sigma^\dagger\chi')\big]^2
- L_8 \left\{
\mathrm{Tr}\big[(\chi'^{\dagger} \Sigma\! +\! \Sigma^\dagger\chi')^2\big]
\!-\! \big[\mathrm{Tr}(\chi'^{\dagger} \Sigma \!+\! \Sigma^\dagger\chi')\big]^2/2
\right\}
\notag \\ &\quad
+ i L_9 \mathrm{Tr}(L_{\mu\nu} D_\mu \Sigma D_\nu \Sigma^\dagger +
           R_{\mu\nu} D_\mu \Sigma^\dagger D_\nu \Sigma)
\notag \\ &\quad
+ \tilde W\; \mathrm{Tr}(D_\mu \Sigma^\dagger D_\mu \Sigma)
         \mathrm{Tr}(\hat{A}^{\dagger} \Sigma + \Sigma^{\dagger}\hat{A})
- W\; \mathrm{Tr}(\chi'^{\dagger} \Sigma + \Sigma^\dagger\chi') 
         \mathrm{Tr}(\hat{A}^{\dagger} \Sigma + \Sigma^{\dagger}\hat{A})
\notag \\&\quad
- W' \big[\mathrm{Tr}(\hat{A}^{\dagger} \Sigma +
       \Sigma^{\dagger}\hat{A})\big]^2 
+ W_{10}\; \mathrm{Tr}(D_\mu \hat{A}^\dagger D_\mu \Sigma + 
          D_\mu \Sigma^\dagger D_\mu \hat{A})
\notag \\&\quad
- H_2 \mathrm{Tr}(\chi'^\dagger\chi')
- H' \mathrm{Tr}(\hat{A}^\dagger \chi'+ \chi'^\dagger \hat{A}) 
\,.
\end{align}
We have dropped the terms proportional to $L_{10}$, $H_1$ and $H_3$,
since these lead only to contact terms in correlation functions,
which we do not need below.
We have also introduced the useful combinations
\begin{equation}
\tilde W = W_{45}-L_{45} \,, \;
W = W_{68} - 2 L_{68} \,, \;
W'= W'_{68} - W_{68} + L_{68} \,, \;
H' = H'_2 - H_2 \,.
\end{equation}
This Lagrangian has the same form as the NLO continuum chiral
Lagrangian aside from the ``$W$'' and $H'$ terms.

We are interested in the terms which are linear in $a$.
Setting aside the $H'$ term, since it contributes
only to condensates, this leaves the terms
multiplied by $\widetilde W$, $W$ and $W_{10}$. The latter
is in fact redundant, as we discuss below, so can be ignored.
The key point is then that, in both the $\widetilde W$
and $W$ terms, the lattice spacing appears in the combination
\begin{equation}
\label{E:Aterm}
\Tr(\hat{A}^\dagger \Sigma + \Sigma^\dagger \hat{A})
\,.
\end{equation}
If we now expand $\Sigma$ about $\Sigma_0$ in the following way,
\begin{equation}
\label{E:LOexpand}
\Sigma = \exp(i \omega_0 \tau_3/2) \exp(i\vec{\pi}\cdot\vec{\tau}/f)
         \exp(i\omega_0 \tau_3/2) \,, 
\end{equation}
then the quantity (\ref{E:Aterm}),
with $\hat{A}$ set to its final value $\hat{a}$,
is invariant under the spurionic symmetry
\begin{equation}
\pi(x) \to - \pi(x)\,, \ 
\omega_0 \to - \omega_0\,,\  
p \to - p\,, \ 
\ell_\mu \leftrightarrow r_\mu
\ \Rightarrow\ \ 
\Sigma \leftrightarrow \Sigma^\dagger\,,\
D_\mu\Sigma \leftrightarrow D_\mu \Sigma^\dagger\,,\
\chi' \leftrightarrow \chi'^\dagger\,.
\end{equation}
It follows that terms even (odd) in pion fields 
must be even (odd) in $\omega_0$. Since the only functions of
$\omega_0$ which can appear are $c_0$ and $s_0$, the former
must be multiplied by an even function of pion fields, the latter
by an odd. Thus at maximal twist ($c_0=0$) the quantity
(\ref{E:Aterm}) produces only odd powers of pion fields,
and, in particular, has no vacuum expectation value.
We also need to consider the combinations
\begin{equation}
\Tr(D_\mu \Sigma^\dagger D_\mu \Sigma)\quad
\textrm{and}\quad
\Tr(\chi'^\dagger \Sigma + \Sigma^\dagger \chi)\,,
\end{equation}
which appear in the $\tilde W$ and $W$ terms.
These are also invariant under the spurionic symmetry, 
but are independent of $\omega_0$,
and thus must be even in pion fields.
Now, when determining the consequences of the NLO terms, it is
sufficient to expand about the LO vacuum, as we have
done in (\ref{E:LOexpand}). It then follows
that the $\widetilde W$ and $W$ terms only give rise, at maximal
twist, to vertices involving odd numbers of pions. The
physical vertices, which involve even numbers of pions, 
do not receive any corrections proportional to $a$ and are thus 
automatically improved.
Of course, the contribution linear in the pion field coming
from the $W$ term leads to an $O(a)$ tadpole, which can convert
LO into NLO vertices with one less pion. However, since the
LO vertices only have even numbers of pions, the resulting
vertices with $O(a)$ corrections all contain an odd number of pions.
Again, the physical vertices are not corrected until $O(a^2)$.

These arguments can also be extended to external sources,
with the conclusion that physical matrix elements are
automatically improved. Conversely, parity violating matrix elements
are maximal at maximal twist.
According to the general considerations of Ref.~\cite{FR03},
these results are expected for $m_q \gg a\Lambda_{\rm QCD}^2$.
Our argument here shows that they hold also in the GSM regime
in which $m_q \sim a\Lambda_{\rm QCD}^2$. This does not require
the use of an improved quark action, but it is essential to
use a definition of maximal twist that sets $m'$, rather than $m$,
to zero. 

A possible concern is how these considerations generalize to
extensions of the chiral Lagrangian incorporating other particles,
e.g. baryons and heavy-light mesons. The point is that,
although the Pauli and mass terms in 
the effective continuum Lagrangian (\ref{E:CLeff}) have 
the same chiral transformation properties,
they are not proportional as operators.
Thus, although they will always enter any extension of
the chiral Lagrangian with
the same form, their relative strength ($B_0/W_0$ in the pionic sector
discussed above) will depend
on the quantity being considered. 
For example, the coefficient of a baryon mass term of the
form $\Tr(\overline B \chi B)$ and that of the corresponding
term caused by discretization errors, $\Tr(\overline B \hat{A} B)$,
will not be in the ratio $B_0/W_0$.
Thus using the shifted mass $m'$ will
not, in general, remove the $\hat{A}$ term. 
It seems, then, that we should be
concerned, when $m\sim a\Lambda_{\rm QCD}^2$, that the
mass and discretization contributions are of the same size in general.
The resolution of this concern is simply that the term linear in $a$
is removed by working at maximal twist.
In fact, we can see this in our calculations, because quantities
such as $f_\pi$, which do not vanish in the chiral limit, have a
similar status in the chiral expansion as does the nucleon mass
(although the details of the expansion differ).

\bigskip

We conclude this section by addressing a few technical points.
First, we discuss the $W_{10}$ term.
As noted above, this term is redundant:
it can be transformed
into a combination of the $W$, $\widetilde{W}$ and $H'$
terms by the change of variables
\begin{equation}
\delta \Sigma = \frac{2 W_{10}}{f^2} 
(\Sigma \hat{A}^\dagger \Sigma -\hat{A}) \,.
\end{equation}
The shifted coefficients of the remaining
three terms are $W+W_{10}/4$, $\widetilde W+W_{10}/2$
and $H'-W_{10}$.
All physical quantities must depend on these combinations
and not on $W_{10}$, $W$, $\widetilde{W}$ and $H'$ 
separately. For this reason, we refer to them as ``physical
combinations'' of LECs, despite the fact that they are
introduced by discretization errors.
Although we could remove $W_{10}$ by this change of variables,
we have kept it, both because we have
found it to provide a useful diagnostic
in the computations of matrix elements, and because it
helps in understanding the impact of improving the underlying
lattice theory, as we now discuss.

We next consider how the effective Lagrangian changes if the
underlying Wilson action, currents and densities are $O(a)$ improved, 
as this illuminates
the nature of the LECs introduced by discretization errors.\footnote{%
We thank the referee for suggesting that we consider the impact
of improvement.}
The required improvement coefficients have been presented in Ref.~\cite{FSW01}.
As already noted, mass dependent $O(a)$ corrections are of higher order
than we consider here. Thus the only improvement coefficients that are
needed are $c_{SW}$, i.e. the coefficient of the lattice Pauli term in the
quark action, and the standard improvement terms for the vector and
axial currents proportional to $c_V$ and $c_A$, respectively. We refer
to Refs.~\cite{FSW01,ALPHA} for details of these terms and discussion of
how they can be determined non-perturbatively. If $c_{SW}$ is set to
its non-perturbative value, then the coefficient $b_1$ in Eq.~(\ref{E:CLeff})
vanishes, so that corrections to the quark effective action are of $O(a^2)$.
It follows that $W_0=0$, and the $O(a)$ shift in $\widetilde m_c$ is absent,
so that $\chi'=\chi$. The chiral effective Lagrangian then takes the same
form as in Eq.~(\ref{E:ChiJLeffpr}), except that the factor of $W_0$
in $\hat{A}$ should be dropped (with concomitant 
changes in the dimensions of various
LECs), and, most importantly, $\widetilde W=W=0$. These latter two
terms, which are proportional to $a \chi$ and $a p^2$ respectively,
are absent because there is no
term linear in $a$ in the underlying quark Lagrangian. The coefficient 
$W_{10}$ does not, however, vanish since it corresponds to $O(a)$ corrections
to currents, and these have not yet been improved. Similarly, $H'$ is non-zero.
Finally, $W'$ is non vanishing since it is proportional to $a^2$.
If the axial current is also non-perturbatively improved,
then one has in addition that $W_{10}=0$ (as can be seen from
the result below for $f_A$). Improving the vector current by adding the
$c_V$ term has no further impact on the LECs, since it turns out
that $c_V$ does not contribute to the quantities we consider at NLO,
as we discuss further below. In summary, if one non-perturbatively improves
the action and currents, then the only remaining discretization errors are
those of $O(a^2)$ proportional to $W'$, as well as the $H'$ term which contributes
terms of $O(a)$ only to the condensate. Our subsequent results show that, in this case, 
while all physical quantities are then $O(a)$ improved, the unphysical parity-flavor
violating matrix elements do have contributions at NLO which are proportional to
$a^2/m_q$ (see sec.~\ref{sec:PV} below).

Finally, we recall that the currents and densities
can be obtained
by taking appropriate functional derivatives of the
action ${\cal S} = \int d^4x \cal L_\chi$: 
\begin{align}
V_\mu^k &= \frac{i}{2} \left(\frac{\delta}{\delta r_\mu^k}
+ \frac{\delta}{\delta \ell_\mu^k} \right) {\cal S}\,,
\label{E:Vbare} \\
A_\mu^k &= \frac{i}{2} \left(\frac{\delta}{\delta r_\mu^k}
- \frac{\delta}{\delta \ell_\mu^k} \right) {\cal S} \,,
\label{E:Abare} \\
S^k &= \frac12  \frac{\delta}{\delta s^k} {\cal S} \,, 
\quad
S^0 =  \frac{\delta}{\delta s^0} {\cal S} \,,
\label{E:Sbare} \\
P^k &=- \frac{i}2  \frac{\delta}{\delta p^k} {\cal S} \,, 
\quad
P^0= -i  \frac{\delta}{\delta p^0} {\cal S} \,, 
\label{E:Pbare} 
\end{align}
where $s = s^0 + \sum_{k=1}^3 s^k \tau_k$,
and similarly for $p$, $\ell_\mu$ and $r_\mu$,
although for the latter two we use only the flavor non-singlet
parts.
Note that these equations give the currents and densities in the
twisted basis, which is that usually used in simulations. 
Our normalizations are such that the corresponding currents and
densities at the
quark level are obtained by taking the same functional
  derivatives as defined above of
\begin{equation}
{\cal S}_{\rm eff} = \int d^4x {\cal L}_{\rm eff} 
- i \bar\psi \gamma_\mu 
\left([r_\mu + \ell_\mu]/2 
                         + \gamma_5 [r_\mu - \ell_\mu]/2\right) \psi
+ \bar\psi (s + i\gamma_5 p) \psi
\end{equation}
Note that the factors of $1/2$ in eqs.~(\ref{E:Vbare}-\ref{E:Pbare})
imply that we are using $\tau_k/2$ in flavor non-singlet operators, 
while for the singlet we do not include the factor of one half.
Thus, for example, $P^k = \bar\psi\gamma_5 \tau_k/2 \psi$
and $S^0 = \bar\psi \psi$ at the quark level.

An important issue is the normalization of the currents.
This can be determined, in the continuum and chiral limits,
by enforcing appropriate Ward identities~\cite{Betal85}.
The currents we use at the quark level automatically satisfy
these identities.
This carries over to the currents in the chiral theory,
since the matching maintains normalizations.
The ratio of the scalar to pseudoscalar densities is also
correctly normalized. The overall normalization of these densities,
however, is a scheme dependent quantity, and this ambiguity is 
reflected in the chiral theory in the presence of the
parameter $B_0$ in $\chi$.
Because of these considerations, the results we give below for
matrix elements of these operators in the effective theory
apply directly to lattice matrix elements, as long
as the lattice operators include their matching factors.

Away from the continuum limit, the explicit breaking
of symmetries implies that one cannot, in general,
normalize the currents and the ratio of densities
in a universal way. 
Different choices of normalization condition will lead
to results differing by $O(a)$, in general.
We stress that this is true both on the lattice
and in the effective chiral theory, as must be the case
since the latter is supposed to represent the former.
This is not, however, an impediment to our calculations.
Since the effective chiral Lagrangian includes all
terms that are allowed by the reduced symmetries of
the lattice theory, our results contain, explicitly,
all possible non-universal terms at the order we work.
Thus in particular our results for matrix elements of
currents and densities should hold for any choice of
lattice operators which are correctly normalized in the
continuum limit.\footnote{
This argument fails, however, if one determines the 
normalization of currents and densities using the
Schr\"odinger functional, as discussed
in the cae of tmLQCD in Ref.~\cite{FSW01}. In that case there are
additional $O(a)$ contributions related to the boundary
fields which are not accounted for by our analysis.
It seems likely that our analysis could be generalized
to include such boundary terms, but we have not
worked this out. Thus we are assuming that the
normalization constants are determined without the use
of boundary fields, e.g. for $Z_V$ and $Z_A$
using the method proposed in Ref.~\cite{Faretal04}.
We thank the referee for pointing out this issue.
}

\section{\label{sec:GSM} Chiral perturbation theory for generic small masses}

In this section we work out in detail the consequences
of the effective chiral Lagrangian in the GSM regime.
In fact, we are now in a position to state precisely how we
define this regime. What we require is that
\begin{equation}
\Lambda_\chi^2 \gg M' \gtrsim \hat{a}
\,,
\end{equation}
where $M'$ is defined in Eq.~(\ref{E:M'def}) and $\Lambda_\chi=4\pi f$.
This ensures that the LO terms in the chiral Lagrangian
(\ref{E:ChiJLeffpr}) dominate over the NLO terms.
This condition can also be written using quark masses:
\begin{equation}
1 \gg \frac{m_q}{\Lambda_{\rm QCD}}
\gtrsim a \Lambda_{\rm QCD}
\,.
\end{equation}
Note that our results for the GSM regime remain
valid if $\hat{a}$ is smaller than $M'$.
The converse does not, however, hold. In particular,
if $M'$ becomes as small as $\hat{a}^2/\Lambda_\chi^2$ then one enters the
Aoki regime where NLO and LO contributions are comparable. 

\subsection{The vacuum and Feynman rules}

The NLO terms cause a small
realignment of the vacuum expectation value of $\Sigma$ away
from $\Sigma_{0}$. In fact, we know from
continuum chiral perturbation theory that the $L_i$ terms
in the potential do not realign the condensate. While this is possible
in principle, given the physical values of the $L_i$,
it does not occur for quark masses in the range of
interest to simulations. Thus
 the only realignment of the condensate is due to the $W$ and
$\widetilde W$ terms.

We define the full NLO condensate to be
\begin{equation}
\langle 0 |\Sigma| 0 \rangle_{NLO} \equiv
\Sigma_{m} \equiv \exp(i \omega_m \tau_3)
= \exp(i [\omega_0 + \epsilon] \tau_3) \,,
\end{equation}
so that $\epsilon=\omega_m-\omega_0$. Minimizing the potential,
we find 
\begin{equation} \label{E:GSMeps}
\epsilon = -\frac{16}{f^2} \hat a s_0 
\left( W + 2 W' \hat{a} c_0/M'\right) \,,
\end{equation}
The $\hat{a}^2/M'$ term is not singular in the GSM regime, since
$M'$ is not allowed to become smaller than $\hat{a}$. We 
note that, to the order we are working,
 we could make the replacements 
$s_0 \to s_m \equiv \sin \omega_m$ and 
$c_0 \to c_m \equiv \cos \omega_m$ in the expression for $\epsilon$.

Later on we will further replace $s_0$ and $c_0$, respectively,
with the sine, $s$,
and cosine, $c$, of a non-perturbatively defined twist angle
$\omega$ which differs from both $\omega_0$ and $\omega_m$ by
$O(a)$.  This is allowed as long as we make the replacement only in NLO
terms. Because of this, from now on
we will use $s$ and $c$ without subscripts in NLO terms,
and point out explicitly where the choice of twist angle is important.

We now expand about the vacuum expectation value, defining
physical pion fields by
\begin{equation}
\label{E:sigmaphys}
\Sigma = \xi_m \Sigma_{ph} \xi_m \,, \quad
\xi_m = \exp(i\omega_m \tau_3/2) \,, \quad
\Sigma_{ph} = \exp(i \vec\pi \cdot \vec\tau / f) \,.
\end{equation}
We use an axial rotation, which has the symmetric form,
since this is the rotation needed at the quark level in
the continuum limit to undo the twist. It gives rise
to pion fields that are in the physical basis, as we will
see below. 

Inserting this expansion into the
chiral Lagrangian (\ref{E:ChiJLeffpr}) we find
\begin{align}
\mathcal{L}_\chi &= 
\frac{\pi^2}{2} (M' + \Delta M') 
- \frac{\pi_3^2}{2} \frac{32}{f^2} \hat{a}^2 s^2 W' \notag \\
&+\frac12 \partial_\mu \vec \pi \cdot \partial_\mu \vec \pi
  \left(1 + \frac{16}{f^2} M' L_{45} 
+ \frac{16}{f^2} \hat{a} c \widetilde W \right) \notag \\
&-\frac{\pi_3}{2} \partial_\mu \vec \pi \cdot \partial_\mu \vec \pi
  \frac{16}{f^3} \hat{a} s \widetilde W
+ \frac{\pi_3 \pi^2}{2} \frac{\epsilon M'}{f} \notag \\
&-(\pi^2)^2 \frac{(M' + 4 \Delta M')}{24 f^2}
+ \pi^2 \pi_3^2 \frac{16}{3 f^4} \hat{a}^2 s^2 W' 
+ \dots \label{E:feynman}
\end{align}
where $\pi^2 = \vec \pi \cdot\vec \pi$, and
\begin{equation}
\Delta M' = \frac{32}{f^2} \left( M'^2 L_{68} 
+ \hat a c M' W + \hat{a}^2 c^2 W' \right) \,.
\end{equation}
We stress again that $s$ and $c$ could equally well be $s_0$ and
$c_0$, or $s_m$ and $c_m$, at NLO accuracy.
With one exception,
we show explicitly in (\ref{E:feynman}) only terms which
we  need for our calculations below.
In particular, we have not included vertices proportional
to $L_1$ and $L_2$, nor four pion vertices with two derivatives.
These are unchanged from continuum chiral perturbation theory [the
rotation in Eq.~(\ref{E:sigmaphys}) cancelling in these terms],
and so lead to NLO contributions of the same form as in the continuum.
The exception are the vertices involving four pions, which we do
not need explicitly below, but which we included to 
show the type of flavor breaking which occurs. 

These results illustrate the generic features discussed in
the previous section.
First, parity and flavor conserving terms of $O(a)$
are also proportional to $c$ and thus vanish at
maximal twist. Second, the $O(a)$ terms which do not vanish at
maximal twist (i.e. the three pion vertices proportional to
$\hat{a} s$) violate parity and flavor. The factor of $s$ ensures that
they vanish when $\mu \to 0$ (when the flavor symmetry is restored). 
Finally, flavor breaking, but parity
conserving contributions (i.e. the additional mass-term for $\pi_3$
and the $\pi^2 \pi_3^2$ vertex) are proportional to $a^2$ and to $s^2
\propto \mu^2$. All these results are expected on general grounds for
$M' \gg \hat{a}$, but also hold, as we see here, for 
$M' \sim \hat{a}$.  

\subsection{Currents and densities in the twisted basis}

At LO, the currents and densities take the usual form
(although other authors use different normalizations):
\begin{align}
V_{\mu,LO}^k &= \frac{f^2}{4} \Tr\left(\tau_k
[\Sigma^\dagger\partial_\mu\Sigma+\Sigma\partial_\mu\Sigma^\dagger]\right)
\,,
\label{E:VLO} \\
A_{\mu,LO}^k &= \frac{f^2}{4} \Tr\left(\tau_k
[\Sigma^\dagger\partial_\mu\Sigma- \Sigma\partial_\mu\Sigma^\dagger]\right)
\,,
\label{E:ALO} \\
S_{LO}^k &= -\frac{f^2 B_0}{4}  \Tr\left(\tau_k
[\Sigma + \Sigma^\dagger]\right) = 0
\,,
\label{E:SaLO} \\
S_{LO}^0 &= -\frac{f^2 2 B_0}{4}  \Tr\left(
\Sigma + \Sigma^\dagger\right) \,,
\label{E:S0LO} \\
P_{LO}^k &= \frac{f^2 B_0}{4}  \Tr\left(\tau_k
[\Sigma - \Sigma^\dagger]\right) \,,
\label{E:PaLO} \\
P_{LO}^0 &= \frac{f^2 B_0}{4}  \Tr\left(
[\Sigma - \Sigma^\dagger]\right) = 0 \,,
\label{E:P0LO} 
\end{align}
where $k = 1,\,2,\,3$. 
The vanishing of the isovector scalar and isoscalar
pseudoscalar densities is a property of $SU(2)$, and holds
also at NLO, so we do not consider these quantities further.

At NLO, the vector and axial currents become
\begin{align}
V_\mu^k &= V_{\mu,LO}^k (1 + {\cal C}) 
+  L_{1},L_2,L_9\ \mathrm{terms} \,,
\label{E:V} \\
A_\mu^k &= A_{\mu,LO}^k (1 + {\cal C}) + 
\frac{4 \hat{a} W_{10}}{B_0 f^2} \partial_\mu P^k_{LO}
+ L_{1},L_2,L_9\ \mathrm{terms} \,,
\label{E:A}
\end{align}
where $k = 1,\,2,\,3$ and
\begin{equation}
{\cal C} = 
\frac{4L_{45}}{f^2} \Tr[\chi'^\dagger\Sigma+\Sigma^\dagger\chi']
- \frac{16\hat{a} \widetilde W}{2 B_0 f^4} S_{LO}^0 \,.
\end{equation}
We do not give the form of the $L_{1}$, $L_2$
and $L_9$ terms since these are
unchanged from the continuum~\cite{GL84}.

For the scalar and pseudoscalar densities at NLO, if we drop
contributions proportional to the sources $s$ or $p$, which give rise
only to contact terms in correlation functions, we find
\begin{align}
S^0 &=  S_{LO}^0 (1 + {\cal D}) - 8 B_0(cM'H_2 + \hat{a} H') \,,
\label{E:S0tw} \\
P^a &= P_{LO}^a (1 + {\cal D}) \,, 
\label{E:Patw} \\
P^3 &=  P_{LO}^3 (1 + {\cal D}) + 4 i B_0 s M' H_2 \,,
\label{E:P3tw}
\end{align}
where $a = 1,\,2$, and
\begin{equation}
\label{E:Ddef}
{\cal D} =  
-\frac{4 L_{45}}{f^2} \Tr[D_\mu \Sigma D_\mu \Sigma^\dagger]
+ \frac{8 L_{68}}{f^2} \Tr(\chi'^\dagger \Sigma +\Sigma^\dagger\chi) 
- \frac{16 \hat{a} W}{2B_0f^2} S_{LO}^0 \,.
\end{equation}
Note that the terms in the chiral Lagrangian proportional to $L_5$,
$L_7$ and $L_8$ give no contribution to these densities,
due to the properties of $SU(2)$ matrices.
The constant terms proportional to $M'$
and $\hat{a}$ in $S^0$ and $P^3$ contribute only their
vacuum expectation values.

We stress that all expressions in this subsection so far
are written in terms of $\Sigma$, whereas to use them we need to
change variables to $\Sigma_{ph}$.
To do this we first need the results:
\begin{align} \label{E:rotVASP}
{V}_{\mu,LO}^{a} &= c_m \hat{V}_{\mu,LO}^a 
                        - \epsilon^{3ab} s_m \hat{A}_{\mu,LO}^b\,,
\quad {V}_{\mu,LO}^3 = \hat{V}_{\mu,LO}^3 \,, \notag \\
{A}_{\mu,LO}^{a} &= c_m \hat{A}_{\mu,LO}^a 
                        - \epsilon^{3ab} s_m \hat{V}_{\mu,LO}^b\,,
\quad {A}_{\mu,LO}^3 = \hat{A}_{\mu,LO}^3 \,, \notag \\
{S}_{LO}^0 &= c_m \hat{S}_{LO}^0 - i s_m 2 \hat{P}_{LO}^3 \,, \quad
{P}_{LO}^3 = c_m \hat{P}_{LO}^3 - i s_m \hat{S}_{LO}^0/2 \,, \quad
{P}_{LO}^a = \hat{P}_{LO}^a \,,
\end{align}
where $a,b=1,2$, and the ``hatted''  currents and densities on the 
right hand sides of these relations 
take the same form as the LO currents and densities in
eqs.~(\ref{E:VLO}-\ref{E:P0LO}) 
but with $\Sigma$ replaced by $\Sigma_{ph}$.
For example
\begin{equation}
\hat{V}_{\mu,LO}^a \equiv \frac{f^2}{4}
\Tr\left(\tau^2[\Sigma_{ph}\partial_\mu\Sigma_{ph}^\dagger
+\Sigma_{ph}^\dagger\partial_\mu\Sigma_{ph}]\right)\,.
\end{equation}
These are LO currents and densities of the chiral effective theory 
expressed in terms of physical pion fields.
The transformations in (\ref{E:rotVASP}) are identical to those
at the quark level between the physical and twisted basis.
We stress that to obtain these results
it is essential to relate $\Sigma$ to $\Sigma_{ph}$
using the axial transformation of Eq.~(\ref{E:sigmaphys}), and not, say,
a left- or right-handed transformation.

Finally, we will also need the results
\begin{align} 
\Tr[\chi'^\dagger\Sigma+\Sigma^\dagger\chi'] &=
M' \Tr[\Sigma_{ph}+\Sigma_{ph}^\dagger] + O(M'\epsilon)
=  - \frac{4 M'}{2 B_0 f^2} \hat{S}_{LO}^0 + O(M'\epsilon)
\label{E:transA}\\
\Tr[D_\mu \Sigma D_\mu \Sigma^\dagger] &=
\Tr[D_\mu \Sigma_{ph} D_\mu \Sigma_{ph}^\dagger]\,,
\label{E:transB}
\end{align}
which allow us 
to express ${\cal C}$ and ${\cal D}$ in terms of
physical pion fields (with the spurion $\chi'$ set to its final
value). We can drop the $O(M'\epsilon)$ term in the first line as
this contributes only at NNLO.

\subsection{Defining the twist angle}

Using the expressions for the currents and densities, we can determine
the twist angle non-perturbatively. In the continuum limit, if the
input twist angle is $\omega=\tan^{-1}(\mu/m)$, the physical currents
and densities are given by 
\begin{align} \label{E:rotating}
\hat{V}_\mu^{a} &= c V_\mu^a + \epsilon^{3ab} s A_\mu^b\,,
\quad \hat{V}_\mu^3 = V_\mu^3 \,, \notag \\
\hat{A}_\mu^{a} &= c A_\mu^a + \epsilon^{3ab} s V_\mu^b\,,
\quad \hat{A}_\mu^3 = A_\mu^3 \,, \notag \\
\hat{S}^0 &= c S^0 + i s 2 P^3 \,, \quad
\hat{P}^3 = c P^3 + i s S^0/2 \,, \quad
\hat{P}^a = P^a \,,
\end{align}
where $a,b = 1,2$, $c = \cos \omega$ and $s = \sin \omega$. 
As already noted, these have the form of the inverse
of the transformations of the LO operators,
Eq.~(\ref{E:rotVASP}), except that here the twist angle
is $\omega$ rather than $\omega_m$.
Note that we can unambiguously create charged pions
using $P^a$ and neutral pions using $A_\mu^3$, since
these operators are invariant.

In the continuum, we can determine $\omega$ by the
condition that there is no flavor or parity breaking in
the physical basis. Different ways of enforcing this 
all lead to the same result for $\omega$. 
On the lattice, however, discretization errors imply that
the different definitions will lead to results differing
by $O(a)$.

We will take as our canonical definition of $\omega$ that
obtained by enforcing 
\begin{equation}
\langle \hat{V}_\mu^2(x) \hat{P}^1(y) \rangle = 0 \,.
\end{equation}
The result for $\omega$ depends, at $O(a)$, on the distance $|x-y|$.
We enforce the relation at long distances, where the single-pion
contribution dominates.
Using the definitions in (\ref{E:rotating}), we can express
the condition as
\begin{equation} \label{E:omegadef}
\tan\omega \equiv \frac{\langle V_\mu^2(x) P^1(y) \rangle}
                       {\langle A_\mu^1(x) P^1(y) \rangle} \,.
\end{equation}
We stress that, at long distances, this gives the same result
as that involving a divergence,
\begin{equation} \label{E:omegadef'}
\tan\omega \equiv \frac{\langle \partial_\mu V_\mu^2(x) P^1(y) \rangle}
                       {\langle \partial_\mu A_\mu^1(x) P^1(y) \rangle} \,.
\end{equation}
since the factors of $\partial_\mu$ acting on the pion propagator
cancel. Both definitions (\ref{E:omegadef}) and
(\ref{E:omegadef'}) have been used in simulations
(see, {\em e.g.}, Refs.~\cite{Faretal04,FarLatt04,Faretal04B}).

To evaluate (\ref{E:omegadef}) at LO we use
the results in Eq.~(\ref{E:rotVASP}). At this order,
only $\hat{A}_{\mu,LO}^1$
and $P^1=\hat{P}^1$ couple to the single pion state,
and we find that $\omega = \omega_m$.
Since $\omega_m=\omega_0$ at LO, we also have that
$\omega=\omega_0$.
This shows that the non-perturbative definition (\ref{E:omegadef})
automatically includes the shift from $m$ to $m'$ discussed above.
In particular, if $\omega=\pi/2$, then $m'=0$ up to corrections
of $O(a^2)$.

The NLO calculation of $\omega$ is simplified by the fact that
most contributions cancel in the ratio
(\ref{E:omegadef}) and do not change the LO result.
This holds for the factors of $1 + {\cal C}$ 
in eqs.~(\ref{E:V},\ref{E:A})
(evaluated with $\Sigma_{ph}=1$),
as well as one-loop wave-function renormalization,
and the one-loop corrections to the coupling of
$\hat{A}_{\mu,LO}^2$ and $P^1$ to the pion.
The $L_1$, $L_2$ and $L_9$ terms do not
contribute at this order. The sole non-trivial
contribution is that from the $W_{10}$ term in $A_\mu^a$, 
and we find 
\begin{equation} \label{E:omegaNLO}
\tan \omega = \frac{s_m}{c_m + \delta} \,, \quad 
\delta = \frac{4 \hat{a} W_{10}}{f^2} \,,
\end{equation}
This result can be rewritten at NLO in a number of useful ways:
\begin{equation}\label{E:omega-omega_m}
s = s_m - \delta s_m c_m \,, \quad c = c_m + \delta s_m^2 \,, \quad 
\omega - \omega_m = - \delta s_m \,.
\end{equation}
Note that in each of these relations one can substitute $s$ for
$s_m$, etc., in the sub-leading term on the right hand side.

As discussed earlier, $W_{10}$ cannot appear alone in a physical
quantity. This is not a concern here, however, since $\omega_m$ is not
physical, i.e. not directly accessible through a non-perturbative
calculation. What must be physical is the difference between results
from alternative non-perturbative definitions of $\omega$.
In particular, we expect $\omega_0$ to be physical since it
contains the physical quark masses. Indeed, the difference
\begin{equation} \label{E:ww0diff}
\omega_0 - \omega =  \frac{16\hat{a}s}{f^2}
\left(W + W_{10}/4 + 2 \hat{a} c W'/M'\right) 
\end{equation}
is given by a physical combination of LECs.

Another non-perturbative definition of $\omega$ can be obtained by
enforcing the continuum relation 
\begin{equation} \label{E:omegadef2}
\langle \hat{V}_\mu^2(x) \hat{A}_\mu^1(y) \rangle = 0 \,.
\end{equation}
This leads to the result
\begin{equation}
\tan(2\omega) =
\frac{\langle A_\mu^1(x) V_\mu^2(y) + V_\mu^2(x) A_\mu^1(y) \rangle}
     {\langle A_\mu^1(x) A_\mu^1(y) - V_\mu^2(x) V_\mu^2(y) \rangle}
\,. 
\end{equation} 
In lattice simulations this relation has been used along with
Eq.~(\ref{E:omegadef}) in order to determine $\omega$ without knowing
the normalization of the vector and axial currents~\cite{Faretal04}. 
Here we do know these normalizations, so this relation gives, in
principle, an independent determination. We find, however, that the
result for $\omega$ is identical to that obtained from
(\ref{E:omegadef}). In fact, this could have been seen in advance
since, when using the single pion contributions, enforcing
both (\ref{E:omegadef2}) and (\ref{E:omegadef}) amounts to
requiring that $\langle 0|\hat{V}_\mu^2|\pi^1\rangle = 0$.

We do obtain a different non-perturbative result for the twist angle 
at non-zero lattice spacing if we enforce
$\langle \hat{S}^0(x) A_\mu^3(y) \rangle = 0$. This leads to 
\begin{equation}\label{E:omegaPdef}
\tan\omega_P \equiv \frac{i \langle S^0(x) A_\mu^3(y)\rangle}
                         {2 \langle P^3(x) A_\mu^3(y)\rangle} \,, 
\end{equation}
where we have denoted the new angle $\omega_P$ (and will use
$c_P = \cos \omega_P$, etc.). A straightforward calculation leads to
\begin{equation}\label{E:omegaPres}
\omega_P - \omega = \frac{4\hat{a} s (4 W + W_{10})}{f^2} 
\end{equation}
at NLO.
The difference between $\omega$ and $\omega_P$ 
contains a physical combination of LECs, as it must since
they both can be calculated non-perturbatively.

Another way of stating this result is to use
$\omega$ obtained from (\ref{E:omegadef})
and calculate the coupling of the
physical scalar density to $\pi_3$. The mismatch of twist angles
implies a non-zero result, which can be expressed as
\begin{equation} \label{E:twmismatch}
\frac{\langle \hat{S}^0(x) A_\mu^3(y)\rangle}
     {\langle \hat{P}^3(x) A_\mu^3(y)\rangle}
= \frac{-8 i \hat{a}s(4W + W_{10})}{f^2} \,,
\end{equation}
where again we take only the single-pion contribution.
The result would, of course, have vanished
if the physical scalar density had been
constructed using $\omega_P$ rather than $\omega$.
We note that (\ref{E:twmismatch}) 
[or (\ref{E:omegaPdef})] provides a method of calculating
one combination of the LECs that are introduced by discretization
errors. We return to this point in more detail below.

In summary, using the currents and densities we have obtained
two non-perturbative definitions of the twist angle. These
are equivalent in the continuum limit, but differ by $O(a)$
away from this limit. This $O(a)$ ambiguity in the twist angle
is maximal at maximal twist. Nevertheless, as we show below,
it affects masses and physical matrix elements only at $O(a^2)$,
and thus does not contradict automatic $O(a)$ improvement.
Only flavor-parity violating quantities are affected at $O(a)$.

\subsection{Constructing the physical currents and densities}

Using the NLO result for $\omega$ defined by (\ref{E:omegadef}), which
we take as canonical from now on, we can now explicitly construct the
physical currents and densities, using the relations in
Eq.~(\ref{E:rotating}). We find for the currents (with $a,b=1,2$): 
\begin{align}
\hat{V}_\mu^a &= \hat{V}_{\mu,LO}^a (1 + {\cal C}) 
- \frac{4 \hat{a} s W_{10}}{f^2} \epsilon^{3ab}\hat{A}_{\mu,LO}^b
+ \frac{4 \hat{a} s W_{10}}{B_0 f^2} 
  \epsilon^{3ab} \partial_\mu \hat{P}_{LO}^b 
+ L_1,L_2,L_9 \ \mathrm{terms} \,,
\label{E:Vaphys}\\
\hat{V}_\mu^3 &= \hat{V}_{\mu,LO}^3 (1 + {\cal C})
+ L_1,L_2,L_9 \ \mathrm{terms} \,,
\label{E:V3phys} \\
\hat{A}_\mu^a &= \hat{A}_{\mu,LO}^a (1 + {\cal C})
-\frac{4 \hat{a} s W_{10}}{f^2}\epsilon^{3ab}\hat{V}_{\mu,LO}^b
+ \frac{4 \hat{a} c W_{10}}{B_0 f^2} \partial_\mu \hat{P}_{LO}^a
+ L_1,L_2,L_9 \ \mathrm{terms} \,,
\label{E:Aaphys}\\
\hat{A}_\mu^3 &= \hat{A}_{\mu,LO}^3 (1 + {\cal C})
+ \frac{4 \hat{a} W_{10}}{2 B_0 f^2} 
\partial_\mu \left( 2 c \hat{P}_{LO}^3 - i s \hat{S}_{LO}^0 \right)
+ L_1,L_2,L_9 \ \mathrm{terms} \,,
\label{E:A3phys}
\end{align}
where ${\cal C}$ is the same as above, but we now write it
using physical fields:
\begin{equation} \label{E:AVcorr}
{\cal C} = \frac{16}{2 B_0 f^4}
\left[ - (M' L_{45}+ \hat{a}c \widetilde{W}) \hat{S}_{LO}^0
       + \hat{a}s \widetilde{W} 2 i \hat{P}_{LO}^3 \right] \,.
\end{equation}
The  $L_1$, $L_2$, and $L_9$ terms take exactly the
same form as in continuum chiral perturbation theory, 
but now expressed in terms of $\Sigma_{ph}$. This is because
they rotate exactly like the LO currents under the axial
rotation. Since we are interested in flavor-parity breaking
contributions we do not give these terms explicitly.

For the densities, the results are 
\begin{align}
\hat{S}^0 &=  \hat{S}_{LO}^0 (1 + {\cal D})
- \frac{4\hat{a} s W_{10}}{f^2} 2i \hat{P}_{LO}^3
- 4i B_0 \hat{a} s H' \,, 
\label{E:S0phys} \\
\hat{P}^a &= \hat{P}_{LO}^a (1 + {\cal D}) \,,
\label{E:Paphys} \\
\hat{P}^3 &=  \hat{P}_{LO}^3 (1 + {\cal D})
- \frac{4\hat{a} s W_{10}}{f^2} \frac{i\hat{S}_{LO}^0}{2}
- 8B_0 (M' H_2 + \hat{a} c H') \,,
\label{E:P3phys}
\end{align}
where we now express ${\cal D}$ in terms of physical fields:
\begin{equation} \label{E:PScorr}
{\cal D} =
-\frac{4 L_{45}}{f^2} \Tr[D_\mu \Sigma_{ph} D_\mu \Sigma_{ph}^\dagger]
+ \frac{16}{2 B_0 f^4}
\left[ - (2 M' L_{68}+ \hat{a}c W) \hat{S}_{LO}^0
       + \hat{a}s W 2 i \hat{P}_{LO}^3 \right] \,.
\end{equation}

For both currents and densities, the mismatch between $\omega$
and $\omega_m$ leads to contributions on the right hand sides
proportional to $W_{10}$.
Note that the currents and densities themselves need not
be composed of physical combinations of LECs, since they are not
directly measurable. It is their matrix elements, such as those
we compute below, which are physical.

One test of the results given above
is that the single pion matrix elements of
$\hat{V}_\mu^a$ should vanish, i.e. the physical vector currents do
not couple to single pions. There are contributions both from the
$\hat{A}_{\mu,LO}^b$ and $\partial_\mu\hat{P}_{LO}^b$ terms and these
do cancel. In effect, our condition for determining $\omega$ has
enforced this result. 

There is a subtlety concerning the choice of twist angle.
The results above for the physical currents and densities hold
as written for our canonical choice, $\omega$. 
If instead we had used $\omega_P$ to define the rotation between
twisted and physical bases, then some of the explicitly parity-flavor
breaking terms are changed. In particular, in the terms proportional
to $\hat{A}_\mu^b$ in Eq.~(\ref{E:Vaphys}),
to $\hat{V}_\mu^b$ in Eq.~(\ref{E:Aaphys}),
to $\hat{P}^3$ in Eq.~(\ref{E:S0phys}),
and to $\hat{S}^0$ in Eq.~(\ref{E:P3phys}),
$W_{10}$ would be replaced by $-4 W$. 
The other terms proportional to $W_{10}$ are
not, however, changed. In our subsequent results,
these changes would only impact the flavor-parity violating
matrix elements.

Finally, we note the same general features in these results as
observed earlier in the Feynman rules. Discretization errors which
do not violate flavor and parity come with factors of $c$, and thus
vanish at maximal twist. Flavor-parity breaking terms, however,
are proportional to $\hat{a} s$.

\subsection{\label{subsec:piM}Pion masses}

With the Feynman rules, currents and densities in hand, we now
turn to the predictions for masses and matrix elements.
We begin with the charged pion mass, which we find at NLO to be
given by
\begin{align}
\label{E:mpi2}
m_{\pi_{1,2}}^2 &= M' + \frac{16}{f^2}\left(
M'^2 (2L_{68}-L_{45}) + M' \hat{a} c (2 W-\widetilde W)
+ 2 \hat{a}^2 c^2 W'\right) + \textrm{1-loop} \,.
\end{align}
Here the one-loop contribution is unchanged from that in the
continuum as long as the result is expressed in terms
of the LO pion mass-squared $M'$.
This is because it involves only LO vertices and masses,
which themselves have the same form as in the continuum.
To be completely clear on this point we quote the result
in this case
\begin{equation}
\textrm{1-loop} = \frac{M'^2}{2 \Lambda_\chi^2}
\ln\left(\frac{M'}{\Lambda_R^2}\right)
\,,
\end{equation}
where $\Lambda_R$ is the renormalization scale.
Since our emphasis is on discretization errors, however,
we will not give explicit expressions for the 1-loop contributions
to other quantities below. They can be found in the original
work on two-flavor ChPT~\cite{GL84}, 
as well as in more recent works including
the extension to the partially quenched theory,
{\em e.g.} Refs.~\cite{RS02,MSch04}.

There are various checks on our result (\ref{E:mpi2}).
When $s\to 0$, it agrees with that of Ref.~\cite{BRS03},
where $m_\pi^2$ was calculated to NLO in the untwisted Wilson theory. 
It goes over to that of Ref.~\cite{MSch04} if we take $\hat{a}\sim M'^2$.
And, after considerable algebra, it can be shown to agree with the
result of Ref.~\cite{Scor04}, where the same quantity was obtained.

Various features of the result are noteworthy. First, it depends
only on physical combinations of LECs, as required. Second,
it is automatically $O(a)$ improved at maximal twist 
($c=0$), or after mass averaging. Finally, the $a^2$ correction
provides a further shift to the critical mass. This shift vanishes,
however, at maximal twist. Thus we find the somewhat surprising
result that, at maximal twist,
 the charged pion mass differs from the continuum
only by terms of NNLO in our expansion.
This is not to say that the result is $O(a^2)$ improved, because
there are contributions proportional to $M' a^2$.

The flavor breaking in the pion masses at NLO is given solely by the
analytic contribution, and we find
(in agreement with Ref.~\cite{Scor04}) that
\begin{align}
\label{E:massplit}
m_{\pi_3}^2 -  m_{\pi_{1,2}}^2 &= -\frac{32}{f^2} \hat{a}^2 s^2 W'
\\ \label{E:massplit2}
&= -W' \frac{32}{f^2} 
  \frac{\hat{a}^2 \hat\mu^2}{(\hat{m}+\hat{a})^2 + \hat{\mu}^2} \,.
\end{align}
To obtain the final form we have used the fact that we can replace
$s$ with $s_0$ at the order we are working.
We know on general grounds that this splitting must vanish 
quadratically in $a \mu$, because it does not violate parity. 
Naively, then, one might have expected the splitting to arise
first at fourth order in our expansion (NNNLO).
What our result shows is that, in fact, there is mass
dependence in the numerator such that the effect is of NLO.

Calculating the pion mass splitting in practice is
complicated by the fact that the neutral pion propagator
includes disconnected quark contractions.
It would nevertheless be an interesting quantity to determine.
For one thing, it would give an
indication of the size of discretization errors. Furthermore,
as noted in Ref.~\cite{Scor04}, determining $W'$ gives information
about the phase diagram in Aoki region. In particular,
if $W'<0$ there is an Aoki phase
on the untwisted Wilson axis, while if $W'>0$ there is a first order
transition on the Wilson axis extending out into the 
twisted plane~\cite{Mun04,Scor04,SW04,SWLatt04}.
It is perhaps surprising that a simulation in the GSM regime 
can give information about the Aoki regime. The reason for
this result is that the same terms in the chiral Lagrangian are
responsible both for the mass splitting and for determining the phase
structure. In fact, the first form of the result, (\ref{E:massplit}),
holds also in the Aoki regime, although the final form,
(\ref{E:massplit2}), does not. 
We discuss this further in the next section.

\subsection{Decay constants and PCAC quark mass}

Using the expressions for the axial current and pseudoscalar
density in the physical basis,
 we can determine their one-pion matrix elements up to NLO:
\begin{align} 
f_A &= f \left\{1 + \frac{4}{f^2}[2 M' L_{45} 
+ \hat a c (2 \tilde W + W_{10})] 
+ \textrm{1-loop} \right\} \,, 
\label{E:fA} \\
f_P &= f B_0 \left\{ 1 + \frac{8}{f^2}\left[
M' (4 L_{68} - L_{45}) + \hat{a}c(2W-\tilde W )\right]
+ \textrm{1-loop} \right\} \,.
\label{E:fP}
\end{align}
Here, as for the pion masses,
the one-loop term is the same as in continuum chiral
perturbation theory~\cite{GL84} and we do not give it explicitly.
At this order there is no flavor breaking---this enters first
at $O(a^2)$, which is NNLO for the decay constants.
The result shows the expected automatic $O(a)$
improvement at maximal twist or under mass averaging, 
and the appearance of a physical combination of LECs. 

In fact, the quantities that we have discussed so far---$\omega$,
$m_\pi^2$, $f_A$ and $f_P$---are not independent.
To see this we first consider the
so-called ``PCAC quark mass'' defined through 
\begin{equation}
m_{PCAC} \equiv 
\frac{\langle \partial_\mu A_\mu^a(x) P^a(y) \rangle}
     {2 \langle P^a(x) P^a(y) \rangle} \,,
\end{equation}
where $a=1,2$. Note that it is not the physical axial current which
appears, but rather that in the twisted basis. 
This quantity is of interest because in the continuum
limit it gives the untwisted component of the quark mass.
We evaluate the correlators for $|x-y|\to\infty$ so that the
single pion contribution dominates.
At LO we find 
\begin{equation}
m_{PCAC}^{LO} = m' = \frac{\mu}{\tan \omega_0} \,,
\end{equation}
showing that this quantity automatically includes the
$O(a)$ offset in the untwisted mass.. 
At NLO we find 
\begin{align}
m_{PCAC} &= \frac{c\, f_A^{NLO} (m_{\pi_a}^2)^{NLO}}{2 f_P^{NLO}} \,,
\label{E:mPCACres} 
\end{align}
where the superscripts indicate that the full NLO
expressions must be used. In this result, we cannot change the
overall factor of $c=\cos\omega$ into, say, $c_m$ or $c_0$.

The result (\ref{E:mPCACres})
also shows that $m_{PCAC}$ is not independent of
the quantities considered so far. In particular, using the
condition $m_{PCAC} = 0$ to determine maximal twist is
equivalent to setting $\tan\omega=\infty$ in (\ref{E:omegadef}). 
Both rely on the vanishing of the coupling of the axial current 
in the twisted basis to the pion.

To proceed further we recall that the lattice symmetry has
an exact ``PCVC'' relation
\begin{equation}
\partial_\mu^\ast \tilde{V}_\mu^k(x) = -2 \mu_0 \epsilon^{3kl} \tilde P^l(x)
\,, \quad k,l = 1,\,2,\,3,\,.
\end{equation}
where $\partial_\mu^\ast$ is the
backward lattice derivative, $\tilde P^k$ the local bare pseudoscalar 
density, and $\tilde V$ the point-split current:
\begin{equation}
\tilde{V}_\mu^k(x) = \frac{1}{2} \Big\{
\bar{\psi}(x)(\gamma_\mu-1)\frac{\tau^k}{2}
 U(x,\mu)\psi(x+a\hat{\mu})
+\bar{\psi}(x+a\hat{\mu})(\gamma_\mu+1)\frac{\tau^k}{2}
 U(x,\mu)^{-1}\psi(x) \Big\} \,.
\end{equation}
The same relation must hold in the effective quark and meson
theories, to all orders in our expansion, if we normalize
our operators to maintain continuum Ward identities. In particular,
at the meson level, we have 
\begin{equation}
\partial_\mu {V}_\mu^k(x) = -2 \mu \epsilon^{3kl} P^l(x)
\,.
\end{equation}
Using this, and the definition of
the twist angle (\ref{E:omegadef}), we find
\begin{equation}
m_{PCAC} =
\frac{\langle \partial_\mu A_\mu^1(x) P^1(y) \rangle}
     {\langle \partial_\mu V_\mu^2(x) P^1(y) \rangle} \times
\frac{\langle \partial_\mu V_\mu^2(x) P^1(y) \rangle}
     {2 \langle P^1(x) P^1(y) \rangle} 
= \frac{\mu}{\tan\omega}
\,,
\end{equation}
This relation should hold to all orders in tm$\chi$PT,
and we have checked that it is valid at NLO.
Using Eq.~(\ref{E:mPCACres}), we see that
$f_A$, $f_P$, $m_\pi^2$ and $\omega$ are related.

\subsection{Parity conserving  matrix elements}

We next consider other flavor-parity conserving matrix
elements that can  be calculated in numerical simulations.
The first example is pion vector form factor,
which can be obtained from the matrix elements
\begin{equation}
\langle \pi_l(p_1) | \hat{V}_\mu^k | \pi_m(p_2) \rangle 
\,.
\end{equation}
We find that, at NLO, the result is unchanged from that in the
continuum, which is given, for example, in Ref.~\cite{GL84}.
This holds true at {\em any} twist angle. It follows because the
contribution from the factor ${\cal C}$ in (\ref{E:AVcorr})
cancels that from wave-function renormalization.

It is perhaps surprising that there are no $O(a)$ terms even
for untwisted Wilson fermions. We know in this case that
to remove $O(a)$ terms in general we need to add to the current
a term containing the tensor bilinear. One
can show, however, that this contributes to the pion matrix
element terms suppressed by $O(a q^2)$ compared to the LO contribution.
While of $O(a)$, these are NNLO corrections in our power counting scheme.

Our next quantity
is the scalar form factor of the pions.
This are flavor conserving at NLO, and can be obtained from
the matrix elements:
\begin{align}
\langle \pi_k(p_1) | \hat{S}^0 | \pi_k(p_2)\rangle
&= 2 B_0 \Bigg\{1 + \frac{8}{f^2}\Big[ - q^2 L_{45}
+ 4 M' (2 L_{68}-L_{45}) + 2 \hat{a} c (2 W-\tilde W) 
\Big] + \textrm{1-loop} \Bigg\} \,.
\end{align}
As usual, the one-loop contribution is the same as in the
continuum~\cite{GL84}. 
Unlike the vector form factor, here $O(a)$ improvement occurs only at
maximal twist.

We consider here also vacuum expectation values of $2 i P^3$ and
$S^0$. The former is the most interesting since
it is calculable on the lattice and, at maximal twist
in the continuum limit, it gives the physical condensate 
$\langle 0|\bar u u + \bar d d |0\rangle$.
It is calculable because $P^3$ does not mix with the identity operator
in the chiral limit. Away from this limit, the mixing is proportional
to $\mu/a^2$ and can, in principle, be
extrapolated away. By contrast, $S^0$ mixes with the identity 
operator for all quark masses, and thus is very
hard to calculate on the lattice.
Nevertheless, we quote results for it
as they illustrate an interesting theoretical point. 

Our results for these condensates at NLO are:
\begin{align}
\langle 2 iP^3 \rangle &= -2 f^2 B_0 s \left\{1 + \frac{4}{f^2}\left[ 
M'(8 L_{68}+H_2) + \hat{a}c(4 W + W_{10}) \right] 
+ \textrm{1-loop} \right\} \,, \\
\langle S^0 \rangle &= -2 f^2 B_0 c \left\{1 + \frac{4}{f^2}\left[
M'(8 L_{68}+H_2) + \hat{a}c(4 W + W_{10})\right] 
+ \textrm{1-loop} \right\} + 8 B_0 \hat{a} (W_{10}-H') \,.
\end{align}
Here the relative 1-loop correction is of the same form for both quantities
and of the same as for the continuum condensate.
We stress that the inclusion of the $H'$ term in the chiral
Lagrangian is essential to make the result for $\langle S^0 \rangle$
physical. We note that $\langle P^3\rangle$ is $O(a)$ improved at
maximal twist, which is expected because it is proportional to the
physical condensate (in the chiral limit). We note also that the $M'$
term in $\langle P^3 \rangle$, since it is multiplied by the overall
factor of $s$, is indeed proportional to $\mu$. Thus $H_2$ is the
constant which contains the quadratic divergence. On the other hand,
$S^0$ is not a multiplicatively renormalizable operator, as it mixes
with the identity operator, even in the chiral limit. Thus the general
considerations of Ref.~\cite{FR03} do not require that it
be automatically $O(a)$ improved, and indeed it is not,
as shown by the last term.

\subsection{Parity violating matrix elements}
\label{sec:PV}

We finally consider simple examples of unphysical, parity violating
quantities which vanish in the continuum due to flavor and parity
symmetries, but are
present in tmLQCD at $O(a)$ since they are not automatically
improved. The first such quantities are axial form factors of the pions.
These are obtained from the matrix elements
\begin{equation}
\langle \pi_a |\hat{A}_\mu^{a}|\pi_3\rangle \,,\
\langle \pi_a |\hat{A}_\mu^{3}|\pi_a\rangle \,,\
\langle \pi_3 |\hat{A}_\mu^{3}|\pi_3\rangle \,,
\end{equation}
where $a=1,2$. Calculating these in lattice simulations is
straightforward in principle as they involve only single particle 
states. 
The main complication is that there are quark-disconnected
contractions in addition to the usual quark-connected contractions. 

\begin{figure}[htbp]
\centering
\subfigure[Direct term]{
\label{fig:feynm:a}
\scalebox{1}[1]{\includegraphics[width=2.5in]{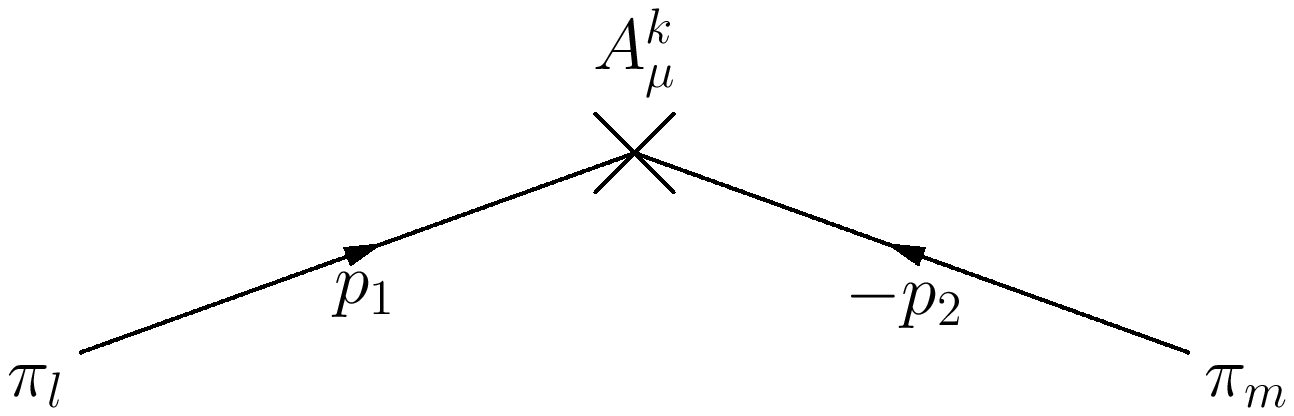}}}
\hspace{0.5in}
\subfigure[Pole term]{
\label{fig:feynm:b}
\scalebox{1}[0.7]{\includegraphics[width=2.5in]{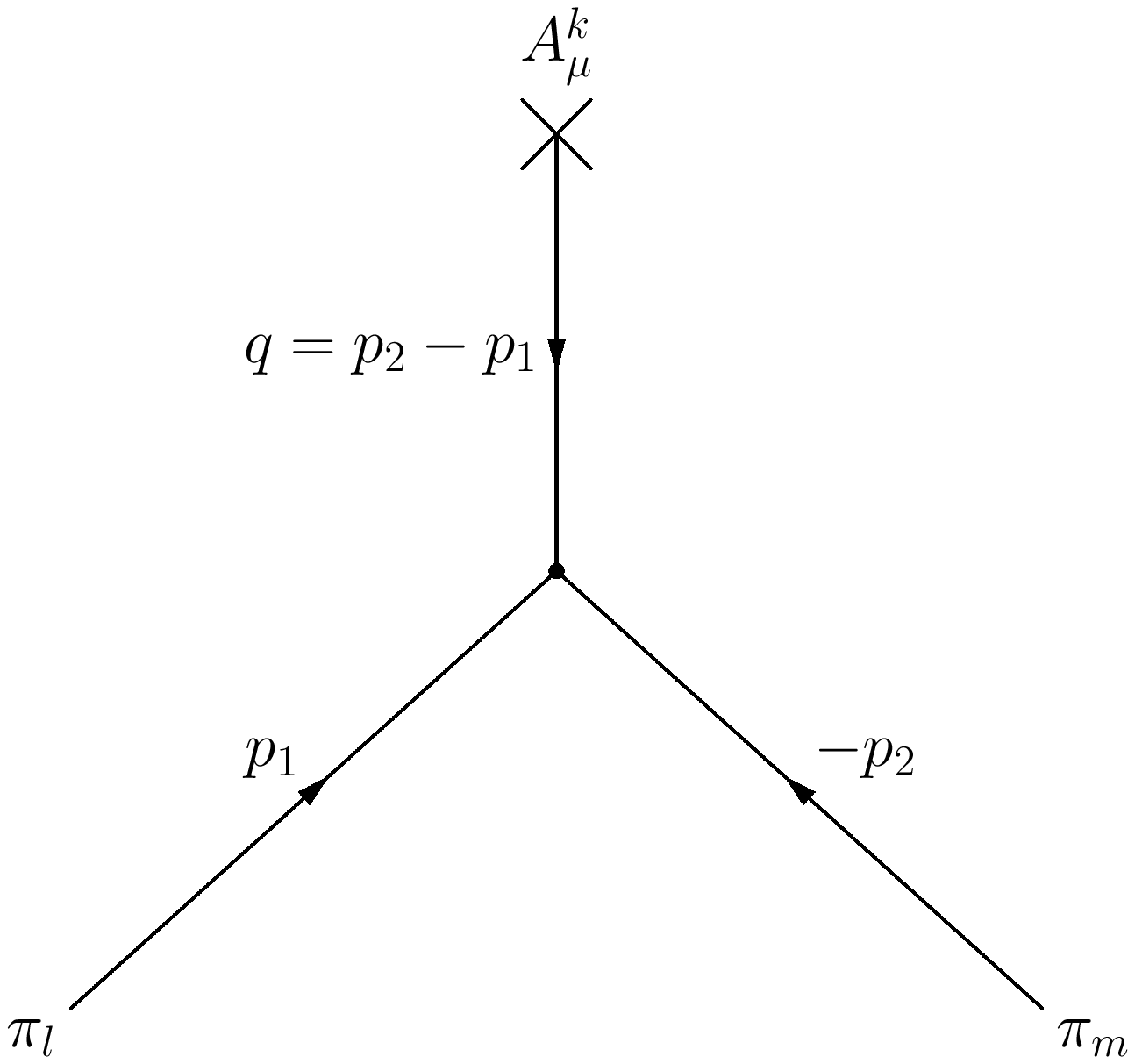}}}
\caption{\label{fig:feyn} Generic types of Feynman diagrams that
  contribute to the two pion matrix elements of the physical axial
  current. The arrows label only the momentum flow. The direct
  term comes from the two pion vertices in the axial current. The
  pole term arises from contracting the single pion term in the axial
  current with the three pion vertices in $\mathcal{L}_\chi$.} 
\end{figure}

At NLO, the contributions to these matrix elements are of two types,
as shown in Fig.~(\ref{fig:feyn}): the direct two pion terms in the
axial current (both the $\hat{A}_{LO}^a \hat{P}_{LO}^3$ and the
$\hat{V}_{LO}^b$ terms), and a pole term in which the axial current
creates a single pion which then connects with the two external pions
through the three pion vertices in $\cal{L}_\chi$. 
There is no 
contribution from the $L_1$, $L_2$, or $L_9$ terms since these do not break
flavor or parity at this order. We also do not need to include
wavefunction renormalization since the leading contribution is of
NLO. There are also no one-loop contributions. Our results are
(with $a=1,\,2$):
\begin{eqnarray}
\langle \pi_a (p_2) |\hat{A}_\mu^{a}| \pi_3(p_1)\rangle &=&
\frac{16 \hat{a} s }{f^2} \times \Bigg\{
(p_1)_\mu
\left[\frac{W_{10}}{4} +  W + \frac{2 \hat{a}c W'}{q^2 + M'}
+  \frac{(\tilde W/2 -W) q^2}{q^2 + M'} \right] 
\nonumber\\&&
\mbox{}\quad - (p_2)_\mu
\left[-\frac{W_{10}}{4} +  W - \widetilde W + \frac{2 \hat{a}c W'}{q^2 + M'}
+ \frac{ (\widetilde W/2-W) q^2}{q^2 + M'} \right]\Bigg\}\,,
\label{E:piaAapi3}
\\
\langle \pi_a(p_2) |\hat{A}_\mu^{3}|\pi_a(p_1)\rangle &=&
\frac{16 \hat{a} s }{f^2} 
(p_1-p_2)_\mu
\left[\frac{-W_{10}}{4} +  W -  \widetilde W
+ \frac{2 \hat{a}c W'}{q^2 + M'}
+  \frac{(\widetilde W/2-W) q^2}{q^2 + M'} \right]\,,
\label{E:piaA3pia}
\\
\langle \pi_3(p_2) |\hat{A}_\mu^{3}|\pi_3(p_1)\rangle &=&
\frac{16 \hat{a} s}{f^2} 
(p_1-p_2)_\mu
\left[-\frac{W_{10}}{4} + 3 W - 2 \widetilde W
+ \frac{6 \hat{a}c W'}{q^2 + M'}
+\frac{3(\widetilde W/2 - W) q^2}{q^2 + M'} \right]
\label{E:pi3A3pi3}
\end{eqnarray}
where $q=p_2-p_1$. A check on the results is that all quantities are
composed of physical combinations of LECs. 
Note that the overall factor is maximized at maximal twist.

The final quantities we consider are the pseudoscalar form factors
of the pions. The required matrix elements
are similar to those for the
axial currents, and have both direct and pole contributions. Our
results are (with $a=1,\,2$):
\begin{eqnarray}
\langle \pi_a (p_2) |\hat{P}^{a}|\pi_3(p_1)\rangle &=&
\frac{16 \hat{a} s i B_0}{f^2} 
\left[\frac{2 \hat{a}c W'}{q^2 + M'}
+  \frac{(\tilde W/2 -W) q^2}{q^2 + M'} \right] 
\label{E:piaPapi3}
\\
\langle \pi_a(p_2) |\hat{P}^{3}|\pi_a(p_1)\rangle &=&
\frac{16 \hat{a} s i B_0}{f^2} 
\left[\frac{-W_{10}}{4} +  W -  \widetilde W
+ \frac{2 \hat{a}c W'}{q^2 + M'}
+  \frac{(\widetilde W/2-W) q^2}{q^2 + M'} \right]\,,
\label{E:piaP3pia}
\\
\langle \pi_3(p_2) |\hat{P}^{3}|\pi_3(p_1)\rangle &=&
\frac{16 \hat{a} s iB_0}{f^2}
\left[-\frac{W_{10}}{4} + W - \widetilde W
+ \frac{6 \hat{a}c W'}{q^2 + M'}
+\frac{3(\widetilde W/2 - W) q^2}{q^2 + M'} \right]
\,.
\label{E:pi3P3pi3}
\end{eqnarray}
These are very closely related to the corresponding axial matrix
elements. 

The results for the parity-flavor violating form factors do depend on
the choice of definition of twist angle.
The results above are for our canonical choice, $\omega$. 
If instead we use $\omega_P$ to rotate to the physical
basis, then we find that in Eqs.~(\ref{E:piaAapi3}),
(\ref{E:piaP3pia}) and (\ref{E:pi3P3pi3})
[though not in Eqs. (\ref{E:piaA3pia}) and (\ref{E:pi3A3pi3})]
the factors of $W_{10}/4$ are replaced by $-W$. 
In particular, this means that all the matrix elements of the
physical pseudoscalar density do not depend on $W_{10}$ with this
definition of twist angle.
This explains a puzzle that appears when considering
the application of our results to an underlying
lattice theory in which the quark action is non-perturbatively improved,
but the currents are not. Since the pseudoscalar density does not
require improvement through NLO in our expansion, one would expect
that its parity violating matrix elements would have no terms
linear in $a$. As discussed above, however, improving the quark
action implies setting $\widetilde W=W=0$ but leaving $W_{10}$ non-zero,
and so the results of Eqs.~(\ref{E:piaP3pia}) and (\ref{E:pi3P3pi3}) 
do not vanish.
The puzzle is explained by noting that the twist angle is
determined using the axial and vector currents, which are unimproved.
If instead one uses the pseudoscalar and scalar densities, i.e.
uses $\omega_P$, then, as just noted, the $W_{10}$ dependence
is absent as expected.

The matrix elements of this subsection provide a way of determining the
LECs  associated with the discretization.
Using the momentum dependence one can, in principle, separately determine
the physical combinations $\widetilde W - 2 W$, $W'$
and $W + W_{10}/4$. Furthermore, since only two of the expressions in
square brackets in (\ref{E:piaAapi3}-\ref{E:pi3A3pi3}) are
independent, and similarly for (\ref{E:piaPapi3}-\ref{E:pi3P3pi3}),
there are a number of predictions implicit in these results.
We stress that these predictions hold only at NLO in tm$\chi$PT,
and will be broken at higher orders.

\subsection{\label{subsec:GSMsumm} Summary of results in GSM regime}

In this final subsection we summarize what we have learnt about the
GSM regime using tm$\chi$PT, and describe one way in which
these results could be used in simulations. We do not discuss the
technical problems that can arise when calculating
the quantities we have discussed  using lattice methods
(e.g. the difficulty in simulating disconnected contractions). We 
simply assume that these difficulties can be overcome. 

We imagine that a simulation is done with particular values of $\mu_0$
and $m_0$, and that all the needed $Z$ factors have been determined.
Thus, given $Z_P$, we know the renormalized twisted mass, $\mu$.
We do not, however, yet know the shifted untwisted mass $m'$;
for this we need to determine the (shifted) critical mass with
$O(a)$ accuracy. Our only assumption about $m'$ is that it (and $\mu$)
are such that we are in the GSM regime.
Our aim is to determine as many as possible of the parameters in
tm$\chi$PT, to make predictions about physical quantities, and
to provide consistency tests of tm$\chi$PT.

One possible way of proceeding is as follows.
\begin{enumerate}
\item
We determine $\omega$ non-perturbatively using Eq.~(\ref{E:omegadef}).
This determines the untwisted quark mass {\em including
the $O(a)$ shift} using
\begin{equation}
m' = \frac{\mu}{\tan\omega_0} = \frac{\mu}{\tan\omega} [1 + O(a)] \,.
\end{equation}
The relative uncertainty of $O(a)$ 
(which corresponds to an absolute uncertainty
of $O(a^2)$ in the GSM regime) means, however, that
the Aoki region cannot be resolved---it collapses to a point at this
accuracy.

\item
To obtain $m'$ to greater accuracy we next calculate $\omega_P$ using
Eq.~(\ref{E:omegaPdef}). From the difference $\omega_P-\omega$, or
equivalently from the ratio given in Eq.~(\ref{E:twmismatch}), we can 
determine $\hat{a}(4 W + W_{10})/f^2$. This is interesting in its own
right, as it not only gives an indication of the size of
discretization errors, but it also allows us to partially ``correct'' 
for the difference between $\omega$ and $\omega_0$. In fact, for
maximal twist ($\omega=\pm \pi/2$) we have, 
using Eqs.~(\ref{E:ww0diff}) and (\ref{E:omegaPres}),
\begin{equation}
\omega_0 = \omega_P \left[1 + O(a^2)\right] 
\qquad \textrm{(Maximal twist)}\,.
\end{equation}
In this case we now know $m'$ with a relative uncertainty of $O(a^2)$,
and an absolute uncertainty of $O(a^3)$.
We stress that this accuracy is to be understood
within our power counting scheme,
in which, for example, shifts in $m'$ proportional to $a\mu^2$,
which are expected to be present, are of $O(a^3)$ and so are
too small to be included.

Because of this result,
one might  argue that it makes more sense to use $\omega_P$
as the canonical choice rather than $\omega$. 
One reason not to do so is that the calculation of $\omega_P$ 
is more difficult as it involves
disconnected quark contractions.

\item
To obtain $\omega_0$ (and thus $m'$) at NLO accuracy
for arbitrary twist angle we next
calculate the pion mass splitting. This indicates 
the size of $a^2$ corrections, and specifically determines the
combination $\hat{a}^2 W'/f^2$. We can then obtain $\omega_0$ 
using
\begin{equation}
\omega_0 = \omega_P - \frac{m_{\pi_3}^2 - m_{\pi_a}^2}{m_\pi^2 \tan\omega}
\,.
\end{equation}
Here $a=1,\,2$, and the mass of any of the pions can be used for
$m_\pi$ in the denominator. 

At this point, we know where our simulation lies in the $m'$, $\mu$ plane
with an $O(a^2)$ relative uncertainty. Thus we know the
physical quark mass $m_q = \mu/s_0 = M'/(2B_0)$ with
similar accuracy. This resolution is good enough to resolve the
Aoki region ($m'\sim\mu\sim a^2\Lambda_{\rm QCD}^3$), 
so we know what parameters to use if we want to study that region.
Indeed, as discussed above, knowing the sign of $W'$ 
we can predict the phase structure in the Aoki region.

\item
We now use one of the parity-violating matrix elements to
determine the final, linearly independent, unknown combination of LECs
associated with discretization errors, namely $2W-\tilde W$. For
example, we could use 
\begin{equation}
 \frac{2 i \langle \pi^a(p_1) |\hat{P}^{3}|\pi^a(p_1)\rangle}
{\langle \pi^a(p_1) |\hat{S}^{0}|\pi^a(p_1)\rangle} =
- \frac{16 \hat{a} s (2 W - \tilde W)}{f^2} 
+ (\omega_P-\omega) + (\omega_P-\omega_0) \,.
\end{equation}
Here we have taken $p_2=p_1$ so that $q^2=0$, and
normalized with the scalar matrix element to remove the
overall factor of $B_0$.

Given this result, we can make absolutely normalized predictions for
the ratios of all the other parity violating matrix elements of
$\hat{P}^k$ to the pion matrix element of $\hat{S}^0$, including their
$q^2$ dependence. Similarly we can predict, including normalization,
the parity violating axial form factors given in
Eqs.~(\ref{E:piaAapi3}-\ref{E:pi3A3pi3}). 
These predictions provide a test
of the applicability of tm$\chi$PT at NLO, as well as an indication of
the size of discretization errors.

\item
Finally, we calculate the physical quantities
$m_{\pi_a}^2$, $f_A$, $f_P$, $\langle P_3 \rangle$ and
$\langle \pi| S, V|\pi\rangle$.
Here we distinguish between maximal and non-maximal twist.
\begin{enumerate}
\item
{\bf Maximal twist}. All these quantities are automatically
$O(a)$ improved, {\em i.e.} they have exactly the same dependence on
$m_q$ as in the continuum up to $O(a^2)$ corrections,
within our power counting scheme.
Our results show that this automatic improvement holds
even if one is working in a region where $m_q \sim a \Lambda_{\rm QCD}^2$.

We stress that this automatic $O(a)$ improvement holds irrespective of
whether we use $\omega$ or $\omega_P$ or some other choice for the
twist angle differing from $\omega$ by $O(a)$.
This is because an $O(a)$ uncertainty in $\omega$ 
leads to an uncertainty in physical quantities 
of size $a^2 s$, so that they are still $O(a)$ improved. 

\item
{\bf Non-maximal twist}. 
Here we find that, having determined the LECs, we can
{\em completely remove $O(a)$ errors in the physical
  quantities that we have calculated}. These errors can simply be
subtracted or divided out. For example, the quantity
\begin{equation}
\left(m^{\rm eff}_{\pi_a}\right)^2 \equiv
\frac{m_{\pi_a}^2 - 32 \hat{a}^2 c^2 W'/f^2}
{1 + 16 \hat{a} c (2 W-\tilde W)/f^2}
\end{equation}
is predicted to have the same dependence on $m_q$
as does $m_\pi^2$ in continuum $\chi$PT, up to $O(a^2)$ errors.

This is an amusing, though perhaps academic result. In practice, when
using tmLQCD, one should clearly work at maximal twist and avoid the
need for such corrections. Nevertheless, it might be worthwhile using
tmLQCD to determine the LECs (which, for small enough quark masses,
are mass independent, though they do depend on the lattice spacing
and gauge action) and then try to correct results 
previously obtained with unimproved untwisted Wilson fermions. 
\end{enumerate}

\end{enumerate}

\section{\label{sec:Aoki} The Aoki regime}

As noted in the introduction, it is of practical interest
to study how the results obtained in the GSM regime change when
one enters the Aoki regime. We recall that the latter is
defined at the quark level as the region where
$m_q\sim a^2 \Lambda_{\rm QCD}^3$, or in tm$\chi$PT
by $M'\sim \hat{a}^2/\Lambda_\chi^2$.

In the Aoki regime, competition between terms of size $M'$ and 
$\hat{a}^2/\Lambda_\chi^2$ leads to a non-trivial phase structure, with
first order transition lines ending at second-order endpoints.
This has been discussed extensively in Refs.~\cite{Mun04,Scor04,SW04}
and we do not recapitulate the analysis here. 
We will present results which hold throughout the Aoki
regime except on the phase transition lines themselves.

Before presenting results we would like to make a general comment on the
concept of $O(a)$ improvement in the Aoki regime. The improvement
program is predicated on the assumption of small discretization
errors. In particular the discretization effects are assumed not to
cause large changes in the vacuum. This assumption clearly breaks down
in the Aoki regime, where, in general,
 the direction of the quark condensate is
determined by a competition between quark masses and discretization
errors and the result differs by angles of $O(1)$ from the continuum
theory. Thus in taking the limit $a \to 0$, quantities do not change
in a manner which is perturbative in $a$, and the language of $O(a)$
improvement cannot be applied. This point has been emphasized
in Ref.~\cite{FrezLatt04}. 

A possible exception to this discussion has been raised in
Ref.~\cite{AB04}. They argue that $O(a)$ improvement at maximal
twist remains valid in the Aoki regime, as long as an
appropriate choice of twist angle is used.
We return to this point at the end of this section.

Irrespective of this general question,
one can simulate in the Aoki regime and 
it is useful to obtain the predictions of tm$\chi$PT.
We stress that, even though the language of $O(a)$ improvement
may not apply, one can still use Symanzik's continuum effective
Lagrangian to study the theory in the Aoki regime.

\bigskip
Now we turn to the results. We will keep the discussion brief,
minimizing technical details, since much of the work is
a straightforward generalization of that in the GSM regime.

As explained in Ref.~\cite{SW04}, we do not need to augment
the chiral Lagrangian used in the GSM regime in order to
study the Aoki regime. Instead, since the power counting is
now $p^2\sim m_q \sim a^2\Lambda_{\rm QCD}^3$, and we work only at
LO in this expansion, we can drop some of the NLO terms used above.
In particular, to study the vacuum alignment and pion masses
we need only keep the $W'$ term---those
proportional to $W$, $\widetilde W$ and the $L_i$
can be dropped. When we consider the currents and densities, however,
a new feature emerges. Since the functional derivatives
defining these quantities effectively ``eat up'' one power of
$p^2$ or $M'$, terms which were of too high order for pion masses
(such as the $W$ term of size $a M'\sim a^3$) now need to be kept.
Indeed, for each quantity that we consider, a careful study is
necessary to determine which terms coming from the GSM regime can be
consistently kept.

As already noted, the fundamental difference between the GSM and Aoki
regimes is that the condensate is no longer closely aligned with the
quark mass. In other words, $\epsilon=\omega_m-\omega_0$ is no longer
of $O(a)$, but instead is generically of $O(1)$. 
An indication of this is that the $W'$ contribution to
$\epsilon$ in Eq.(\ref{E:GSMeps}) is proportional to
$\hat{a}^2/(f^2 M')$ and thus of $O(1)$ in the Aoki regime. 
The precise alignment of the
condensate is determined by a quartic equation which has been given in
various forms in Refs.~\cite{Mun04,Scor04,SW04,AB04}. 
It can be written:
\begin{equation} \label{E:epsAoki}
M' \sin(\omega_m-\omega_0) = M' \sin\epsilon
= -\frac{32 \hat{a}^2 s_m c_m W'}{f^2} + O(a^3) \,.
\end{equation}
Note that one solution is $\omega_m=\omega_0=\pi/2$, so that 
$\epsilon=0$ at maximal twist. The corrections to this
result are of $O(a)$, but we do not control these since they include
contributions from $a^3$ terms in the chiral Lagrangian.

We now work our way through the quantities considered in
the previous section, concentrating on how they differ
in the Aoki regime.
The expressions for the currents and densities,
(\ref{E:A}-\ref{E:Ddef}), are unchanged, except that
the contributions proportional to $M'$ or $p^2$ can be dropped.
We do, however, control the $O(a)$ corrections proportional
to $W$, $\tilde W$ and $W_{10}$.
The change of variables in Eqs.~(\ref{E:rotVASP}) 
and (\ref{E:transA}) also goes through unchanged,
but that in (\ref{E:transB}) is replaced by 
\begin{equation}
\label{E:transB'}
\Tr[\chi'^\dagger\Sigma+\Sigma^\dagger\chi'] =
M'\cos\epsilon \Tr[\Sigma_{ph}+\Sigma_{ph}^\dagger] +
i M'\sin\epsilon \Tr[\tau_3(\Sigma_{ph}-\Sigma_{ph}^\dagger)] \,.
\end{equation}
In fact, this change only matters if this term contributes
at LO, and thus only affects the pion masses, which are
discussed below. For all other quantities the $M'$
contributions are of higher order than we control in the Aoki
regime.

Because of these considerations, the determination of the
twist angles $\omega$ and $\omega_P$ goes through unchanged.
They differ from each other, and from $\omega_m$, by $O(a)$,
with the differences given in Eqs.~(\ref{E:omega-omega_m})
and (\ref{E:omegaPres}). The expressions for the physical
currents and densities, Eqs.~(\ref{E:Vaphys}-\ref{E:PScorr}),
thus remain valid, except that the terms proportional to
$M'$ and $p^2$ can be dropped.
It then follows that the GSM expressions
for $f_A$, $f_P$, the parity conserving matrix elements
and the condensates remain valid in the Aoki regime,
except that we should drop the $M'$ and 1-loop contributions.
We collect the main results here for clarity
\begin{align} 
f_A &= f \left\{1 + \frac{4}{f^2} \hat a c (2 \tilde W + W_{10})
+ O(a^2) \right\} \,, 
\label{E:fAAoki} \\
f_P &= f B_0 \left\{ 1 + \frac{8}{f^2}\hat{a}c(2W-\tilde W )
+ O(a^2) \right\} \,,
\label{E:fPAoki} \\
\langle \pi_k(p_1) | \hat{S}^0 | \pi_k(p_2)\rangle
&= 2 B_0 \left\{1 + \frac{8}{f^2} 2 \hat{a} c (2 W-\tilde W) + O(a^2)
  \right\} \,,
\label{E:FSAoki} \\
\langle 0| 2 iP^3|0 \rangle &= -2 f^2 B_0 s 
\left\{1 + \frac{4}{f^2} \hat{a}c(4 W + W_{10}) + O(a^2)
 \right\} \,.
\label{E:condP3Aoki}
\end{align}
As in the GSM regime, we can replace $s$ and $c$ with $s_P$ and $c_P$,
respectively, to the accuracy we work.
The vector form factor, which we do not display, is simply
given by the leading order term in continuum $\chi$PT with
no $O(a)$ correction.

The results for the pion masses are changed more substantially.
These have already been calculated in Refs.~\cite{Scor04,SW04,AB04}
so we only comment on the relation to our results in the GSM regime.
In the Aoki regime, we need only keep the LO mass term
and the $W'$ term from (\ref{E:ChiJLeffpr}). For the former,
Eq.~(\ref{E:transB'}) shows that 
the contribution to $m_\pi^2$ changes from $M'$ to
$M'\cos\epsilon$ in the Aoki regime.
Combining this with the $W'$ contribution, which is unchanged
from the GSM regime, we find
(in agreement with Refs.~\cite{Scor04,SW04,AB04})
\begin{equation}
\label{E:mpi2Aoki}
m_{\pi_{1,2}}^2 = M'\cos\epsilon  
+ \frac{32 \hat{a}^2 c^2 W'}{f^2} + O(a^3) 
= \frac{\hat{\mu}}{s_m} + O(a^3) 
\,.
\end{equation}
We have used Eq.~(\ref{E:epsAoki}) to obtain the second form.

The mass-squared splitting comes only from $W'$ term
in the chiral Lagrangian. It is easy to see that its
contribution takes the same form as in Eq.~(\ref{E:feynman}),
with $s$ being $s_m$. But since $s$ and $s_m$
differ only by $O(a)$, one can use either at our accuracy.
Thus the result (\ref{E:massplit}) from the GSM regime remains valid
in the Aoki regime, in agreement with Refs.~\cite{Scor04,SW04}.

The other quantities which are substantially changed in the
Aoki regime are the parity violating matrix elements.
These are $O(a)$ in the GSM regime, but become of $O(1)$ in
the Aoki regime. This is indicated by the factors of
$\hat{a}^2/M'$ in the $W'$ contributions in
Eqs.~(\ref{E:piaAapi3}-\ref{E:pi3P3pi3}).
In fact, it is straightforward to see that we control
these quantities only at $O(1)$, and not at $O(a)$.
This simplifies the calculation since only the pole terms
give $O(1)$ contributions. The results are (with $a=1,\,2$): 
\begin{align}
\langle \pi_a(p_2) | \hat{A}_\mu^{a} | \pi_3(p_1)\rangle &=
 q_\mu \frac{M'\sin\epsilon}{q^2+m_{\pi_a}^2} + O(a)
\,,
\\
\langle \pi_a(p_2) | \hat{A}_\mu^{3} | \pi_a(p_1)\rangle &= 
\frac13 \langle \pi_3(p_2) | \hat{A}_\mu^{3} | \pi_3(p_1)\rangle =
 q_\mu \frac{M'\sin\epsilon}{q^2+m_{\pi_3}^2} + O(a)
\,,
\\
\langle \pi_a(p_2) | \hat{P}^{a} | \pi_3(p_1)\rangle &=
- iB_0\frac{M'\sin\epsilon}{q^2+m_{\pi_a}^2} + O(a)
\,,
\\
\langle \pi_a(p_2) | \hat{P}^{3} | \pi_a(p_1)\rangle &=
\frac13
\langle \pi_3(p_2) | \hat{P}^{3} | \pi_3(p_1)\rangle = 
- iB_0\frac{M'\sin\epsilon}{q^2+m_{\pi_3}^2} + O(a)
\,.
\end{align}
We have used Eq.~(\ref{E:epsAoki}) to simplify the expressions.
Note that the choice of pion mass in the denominators makes
an $O(1)$ difference to these quantities.

The expressions in this section all simplify at maximal twist.
As already noted, we would then have $\epsilon=0$ in addition to $c=0$
and $s=1$.
Thus the $O(a)$ corrections to the physical quantities 
in (\ref{E:fAAoki}-\ref{E:condP3Aoki}) vanish, 
so that all are given simply by their LO continuum $\chi$PT values.
The charged pion masses are similarly given by the LO continuum
result, $m_{\pi_a}^2=M'$. The pion mass-squared splitting
does differ from the continuum, but this difference is of $O(a^2)$.
Finally, the parity violating matrix elements become of $O(a)$
again, as in the GSM regime, although we cannot predict them
at the order we are working.

\bigskip

We now return to the issue raised in Ref.~\cite{AB04}.
The authors argue that automatic $O(a)$
improvement at maximal twist does extend into the Aoki regime,
but only if one uses a twist angle based on
a critical mass which includes the $O(a)$ offset, $m'-m$.
They point out that one of the definitions of twist angle 
that accomplishes this is what we call $\omega$ as defined by
  Eq.~(\ref{E:omegadef}) (and which they call $\omega_{WT}$).
We fully agree on the need to use such a definition,
not only in the Aoki regime, but also in the GSM regime,
as we discussed in Sec.~\ref{sec:EChiL}.

As for automatic $O(a)$ improvement, our results support
the proposal of Ref.~\cite{AB04}, although with one caveat.
We argued that, in the Aoki regime, the condensate did not
vary smoothly as the masses are varied. This is true in general,
but not at maximal twist, where the condensate remains fixed,
with $\omega_m=\pi/2$ within an uncertainty of $O(a)$.
Indeed, as we have already noted, 
at maximal twist physical
quantities are described in the Aoki phase
by continuum $\chi$PT up to $O(a^2)$.
If we start in the Aoki regime
at $\omega=\pi/2$, and remain on the $\mu$ axis as $a$ is
reduced (so that we move into the GSM regime and ultimately
to the continuum limit) we expect that physical quantities,
at fixed quark masses, will extrapolate smoothly to the
continuum with errors quadratic in $a$.

The caveat is that the argument will fail if, as $\mu$ is
reduced, one encounters a phase boundary. This is expected
to happen in one of the two possible scenarios for
the phase diagram~\cite{Mun04,Scor04,SW04}. 
In this scenario, in which $W'>0$,
the end-point of the phase boundary
occurs at $\hat{\mu} = 32 \hat{a}^2 W'/f^2$, 
where $m_{\pi_3}=0$.
Values of $\hat{\mu}$ greater than this are allowed,
but for smaller values the condensate changes rapidly
(by an amount of $O(1)$ while the quark mass changes by
$O(a^2)$), and continuum $\chi$PT expressions fail.
In the other scenario ($W'<0$) we see no need to impose
a lower limit on $\mu$---unless $|W'|$ is unnaturally small,
yet higher order terms in the chiral potential will not
change the phase structure.

It is worth noting the accuracy with which the critical mass
must be determined to stay at $\omega=\pi/2$ within an error of
$O(a)$. This is the accuracy that is required to automatically
remove the $O(a)$ terms.
 In the GSM regime, where $\mu\sim a\Lambda_{\rm QCD}^2$,
this requires knowing $m'$ to 
an absolute accuracy of $O(a^2)$. In the Aoki regime, however,
the required absolute accuracy decreases to $O(a^3)$.
This sounds like a difficult goal, but, in fact, as we
discussed in the summary of the previous section, it is attainable
even by doing simulations in the GSM regime.

\section{\label{sec:conc} Conclusion}

In this paper we have used effective field theory methods to
study the discretization errors in tmLQCD. We have presented
results for a number of pionic quantities that can be calculated
in lattice simulations, and studied different possible
definitions of the twist angle. Perhaps the most interesting
quantities to calculate are the difference $\omega-\omega_P$
(which indicates the size of the $O(a)$ uncertainty in the
twist angle), the splitting $m_{\pi_3}^2-m_{\pi_a}^2$ (which indicates the
size of $O(a^2)$ flavor breaking, but parity conserving, 
quantities) and the axial and pseudoscalar form factors of
the pion (which are the simplest examples of unphysical
quantities). To our knowledge, none of these quantities
have been calculated to date, and we urge that they be
considered in the future.

In the introduction, we raised a number of issues concerning
tmLQCD; now we can comment on what we have learnt about these.
Concerning the impact of flavor and parity breaking, we have seen
how this contributes to unphysical quantities at $O(a)$,
and in one case, to physical quantities at $O(a^2)$.
This case is the pion mass splitting, where we confirm the interesting
result of Ref.~\cite{Scor04} that knowledge of the sign of the
splitting allows one to predict the nature of the phase structure
in the Aoki region.
Nevertheless, at NLO in tm$\chi$PT, the pions in the loops are
degenerate, and there is no flavor breaking in decay constants
or vector and scalar form factors. For staggered fermions,
where the ``taste'' breaking between pions is also of $O(a^2)$,
it has been found essential to include this breaking in loop
contributions in order to make good fits to the chiral 
behavior~\cite{MILC}. If this turns out to be true also for tmLQCD
(which depends on the size of the as yet unmeasured flavor breaking),
then the $\chi$PT calculations will need to be extended beyond
NLO. One possibility which we are investigating
is to use a power counting like that in Ref.~\cite{Aoki03}, so
that a full NNLO calculation is not needed.

We also asked which quantities are good indicators of the
size of discretization errors. One answer is that it may be
easier to use the flavor-parity violating quantities
(i.e. the difference between different definitions of twist
angle, and the axial and pseudoscalar form factors),
since these are of $O(a)$ 
(even at maximal twist) and thus easier to calculate.
In more general terms, studying these quantities in simulations
would allow a more thorough  test of our
understanding of tmLQCD.

Another interesting question is
how we can test the reliability of tm$\chi$PT
at the order we are working.
Our results, in fact, contain a number of predictions
that are valid at NLO but not at higher order.
In particular, as we have outlined in the summary subsection of
Sec.~\ref{sec:GSM}, most of the flavor-parity violating form factors
can be predicted once the other quantities that we have discussed have
been calculated. Such tests are important for establishing the
credibility of the chiral and continuum extrapolations that must
ultimately be done. 

We have discussed extensively the issue of the
smallest quark mass that can be used without invalidating
automatic $O(a)$ improvement at maximal twist.
We have argued that one can certainly work in the GSM regime
($m_q \sim a\Lambda_{\rm QCD}^2$)
and also at least part-way into the Aoki regime
($m_q \sim a^2\Lambda_{\rm QCD}^3$).
This is only true, however, if one uses the appropriate definition
of twist angle. Allowable choices are that determined from the
vanishing of the coupling of the bare axial current in the
twisted basis to the charged pion, or from the vanishing of the coupling
of the pseudoscalar density to the neutral pion.
Alternatively one could fix the twisted mass $\mu$
and vary the untwisted mass until the pion masses (either charged
or neutral) reach their minimum values~\cite{AB04}.
What is potentially problematic is to determine the critical quark mass
by extrapolating the squared pion mass to zero along the
untwisted Wilson axis using only moderately small quark
masses. In the GSM regime this may not, in practice,
give the critical mass with the accuracy needed (errors of absolute
size $O(a^2)$) to ensure $O(a)$ improvement at maximal twist.\footnote{%
This point may explain the ``bending phenomenon'' observed
at masses below $m_q\sim a\Lambda_{\rm QCD}^2$ in Ref.~\cite{Biet04}.
Such bending is expected in most quantities if one approaches 
the untwisted Wilson axis along a line parallel to the twisted mass axis.
It would be interesting to fit the results from this work
under this hypothesis, using the formulae we have provided.}
Furthermore, this method does not apply in the Aoki regime, 
where it gives a determination of the critical mass with 
an absolute error of $O(a^2)$, whereas the required accuracy is $O(a^3)$.

One spin-off from our calculation is a method for
determining the new LECs that enter when one
incorporates discretization errors. This both allows a direct measure
of their size, 
and can be used {\em a posteriori} to correct results obtained
with untwisted Wilson fermions.
Furthermore, our calculations extend the results that are
available from $\chi$PT applied to untwisted Wilson 
quarks~\cite{RS02,BRS03} to several new quantities.

Many of the quantities we have considered will be difficult
to calculate in lattice simulations because they involve
quark-disconnected contractions. An important practical question to
consider is whether, by using partially quenched tmLQCD
and its corresponding $\chi$PT~\cite{MSS04},
 one can make predictions for
the quark-connected and disconnected contributions separately.
It will also be interesting to extend the calculations
to the case of a non-degenerate doublet of quarks~\cite{FR03nondegen}.

\section*{Acknowledgements}
We thank Oliver B\"ar and
Chris Sachrajda for correspondence and comments on the manuscript.
SS thanks the School of Physics and Astronomy
at the University of Southampton for its hospitality, which greatly
aided the completion of this work.
This research was supported in part by PPARC grant 
PPA/V/S/2003/00006 and
the U.S. Department of Energy Grant No. DE-FG02-96ER40956.


\begin{thebibliography}{99}


\bibitem{FetalLatt99}
R.~Frezzotti, P.~A.~Grassi, S.~Sint and P.~Weisz,
Nucl.\ Phys.\ Proc.\ Suppl.\  {\bf 83}, 941 (2000).

\bibitem{Fetal01}
R.~Frezzotti, P.~A.~Grassi, S.~Sint and P.~Weisz  [Alpha collaboration],
JHEP {\bf 0108}, 058 (2001).

\bibitem{FrezLatt04}
R.~Frezzotti, 
``Twisted mass lattice QCD'',
plenary talk at Lattice 2004,
hep-lat/0409138.

\bibitem{Kenn04}
A.~D.~Kennedy,
``Algorithms for lattice QCD with dynamical fermions,''
plenary talk at Lattice 2004,
hep-lat/0409167.

\bibitem{FR03}
R.~Frezzotti and G.~C.~Rossi, JHEP {\bf 0408}, 007 (2004)

\bibitem{Pena04}
C.~Pena, S.~Sint and A.~Vladikas,
JHEP {\bf 0409}, 069 (2004).

\bibitem{FR04}
R.~Frezzotti and G.~C.~Rossi,
JHEP {\bf 0410}, 070 (2004).

\bibitem{Faretal04}
F.~Farchioni {\it et al.}, hep-lat/0406039.

\bibitem{FarLatt04}
F.~Farchioni {\it et al.}, hep-lat/0409098.

\bibitem{Biet04}
W.~Bietenholz {\it et al.},
hep-lat/0411001.

\bibitem{Faretal04B}
F.~Farchioni {\it et al.}, hep-lat/0410031.

\bibitem{SS98}
S.~Sharpe and R.~Singleton, Jr, Phys. Rev. D {\bf 58}, 074501 (1998).

\bibitem{Mun04}
G.~M\"unster, JHEP {\bf 0409}, 035 (2004).

\bibitem{Scor04}
L.~Scorzato,
Eur.\ Phys.\ J.\ C {\bf 37}, 445 (2004).

\bibitem{SW04}
S.~R.~Sharpe and J.~M.~S.~Wu,
Phys.\ Rev.\ D {\bf 70}, 094029 (2004).

\bibitem{MSch04}
G.~M\"unster and C.~Schmidt, Europhys. Lett. {\bf 66}, 652 (2004).

\bibitem{RS02}
G.~Rupak and N.~Shoresh,
Phys.\ Rev.\ D {\bf 66}, 054503 (2002).

\bibitem{BRS03}
O.~B\"ar, G.~Rupak and N.~Shoresh, 
Phys.\ Rev.\ D {\bf 70}, 034508 (2004).

\bibitem{SWLatt04}
S.~Sharpe and J.~Wu, hep-lat/0407035.

\bibitem{MunLatt04}
G.~Munster, C.~Schmidt and E.~E.~Scholz,
hep-lat/0409066.

\bibitem{AB04}
S.~Aoki and O.~Bar,
Phys.\ Rev.\ D {\bf 70}, 116011 (2004).

\bibitem{Hash04}
S.~Hashimoto,
hep-ph/0411126.

\bibitem{Aoki}
S.~Aoki,
Phys.\ Rev.\ D {\bf 30}, 2653 (1984);
Phys.\ Rev.\ Lett.\  {\bf 57}, 3136 (1986);
Prog.\ Theor.\ Phys.\ {\bf 122}, 179 (1996).

\bibitem{Sym83}
K.~Symanzik, Nucl. Phys. {\bf B226}, 187 (1983a); 
{\bf B227}, 205 (1983b). 

\bibitem{FSW01}
R.~Frezzotti, S.~Sint and P.~Weisz  [ALPHA collaboration],
JHEP {\bf 0107}, 048 (2001).

\bibitem{WZW}
J.~Wess and B.~Zumino,
Phys.\ Lett.\ B {\bf 37}, 95 (1971); \\
E.~Witten,
Nucl.\ Phys.\ B {\bf 223}, 422 (1983).

\bibitem{GL84}
J.~Gasser and H.~Leutwyler, Annals Phys. {\bf 158}, 142 (1984).

\bibitem{ALPHA}
M.~L\"uscher, S.~Sint, R.~Sommer and P.~Weisz,
Nucl.\ Phys.\ B {\bf 478}, 365 (1996).

\bibitem{Betal85}
M.~Bochicchio, L.~Maiani, G.~Martinelli, G.~C.~Rossi and M.~Testa,
Nucl.\ Phys.\ B {\bf 262}, 331 (1985). 

\bibitem{MILC}
C.~Aubin {\it et al.}  [MILC Collaboration],
Phys.\ Rev.\ D {\bf 70}, 114501 (2004).

\bibitem{Aoki03}
S.~Aoki, Phys. Rev. D {\bf 68}, 054508 (2003).

\bibitem{MSS04}
G.~Munster, C.~Schmidt and E.~E.~Scholz,
Europhys.\ Lett.\  {\bf 86}, 639 (2004).

\bibitem{FR03nondegen}
R.~Frezzotti and G.~C.~Rossi,
Nucl.\ Phys.\ Proc.\ Suppl.\  {\bf 128}, 193 (2004),
hep-lat/0311008.

\end{thebibliography}
\end{document}